\documentclass[preprint]{aastex}

\newcommand{\decsec}[2]{$#1\mbox{$''\mskip-7.6mu.\,$}#2$}
\slugcomment{Accepted for publication in {\it The Astronomical Journal}, Vol. 124, Sep. 2002}

\shorttitle{Margon et al.}
\shortauthors{Faint High Latitude Carbon Stars}

\begin{document}


\title{Faint High Latitude Carbon Stars Discovered by the Sloan Digital Sky
Survey: 
Methods and Initial Results}


\author{Bruce Margon\altaffilmark{1,2}, Scott F. Anderson\altaffilmark{2},
Hugh C. Harris\altaffilmark{3}, Michael A. Strauss\altaffilmark{4}, 
G.~R.~Knapp\altaffilmark{4}, Xiaohui Fan\altaffilmark{5}, Donald P.
Schneider\altaffilmark{6}, Daniel E. Vanden Berk\altaffilmark{7}, David~J.~Schlegel\altaffilmark{4}, Eric~W.~Deutsch\altaffilmark{2,8}, 
\v{Z}eljko Ivezi\'c\altaffilmark{4}, Patrick~B.~Hall\altaffilmark{4,9},
Benjamin~F.~Williams\altaffilmark{2},  
Arthur~F.~Davidsen\altaffilmark{10,11}, J. Brinkmann\altaffilmark{12}, 
Istv\'an Csabai\altaffilmark{13}, 
Jeffrey J. E. Hayes\altaffilmark{12}, Greg~Hennessy\altaffilmark{14}, Ellyne K. Kinney\altaffilmark{12}, S. J. Kleinman\altaffilmark{12},  
Don Q. Lamb\altaffilmark{15},  Dan Long\altaffilmark{12}, Eric~H.~Neilsen\altaffilmark{7},
Robert Nichol\altaffilmark{16}, Atsuko Nitta\altaffilmark{12}, Stephanie A. Snedden\altaffilmark{12}, Donald G. York\altaffilmark{15}
}

\altaffiltext{1}{Space Telescope Science Institute, 3700 San Martin Drive,
Baltimore, MD 21218}
\altaffiltext{2}{Astronomy Department, University of Washington, Box 351580,
Seattle, WA 98195-1580}
\altaffiltext{3}{U. S. Naval Observatory, Flagstaff Station, P.O. Box 1149,
Flagstaff, AZ 86002-1149}
\altaffiltext{4}{Princeton University Observatory, Princeton, NJ 08544-1001}
\altaffiltext{5}{Institute for Advanced Study, Princeton, NJ 08540}
\altaffiltext{6}{Department of Astronomy and Astrophysics, The Pennsylvania
State University, \hbox{University} Park, PA 16802}
\altaffiltext{7}{Fermi National Accelerator Laboratory, P. O. Box 500,
Batavia, IL 60510}
\altaffiltext{8}{Institute for Systems Biology, 1441 N. 34th St, 
Seattle, WA 98103-8904}
\altaffiltext{9}{Pontificia Universidad Cat\'{o}lica de Chile, Departamento de Astronom\'{\i}a
y Astrof\'{\i}sica, Facultad de F\'{\i}sica, Casilla 306, Santiago 22, Chile}
\altaffiltext{10}{Department of Physics and Astronomy, The Johns Hopkins
University, Baltimore, MD 21218}
\altaffiltext{11}{Deceased}
\altaffiltext{12} {Apache Point Observatory, P. O. Box 59, Sunspot, NM 88349-0059}
\altaffiltext{13} {Department of Physics of Complex Systems,
E\"otv\"os University, P\'azm\'ay P\'eter \hbox{s\'et\'any 1/A,}
H-1117, Budapest, Hungary}
\altaffiltext{14} {US Naval Observatory, 3450 Massachusetts Avenue NW,
Washington, DC 20392-5420}
\altaffiltext{15} {Astronomy and Astrophysics Center, University of
Chicago, 5640 South Ellis Avenue, Chicago, IL 60637}
\altaffiltext{16} {Dept. of Physics, Carnegie Mellon University,
5000~Forbes Ave., Pittsburgh, PA~15232}


\begin{abstract}

We report the discovery of 39 Faint High Latitude Carbon Stars (FHLCs)
from Sloan Digital Sky Survey commissioning data. The objects,
each selected photometrically and verified spectroscopically, range
over $16.6 < r^* < 20.0$, and show a diversity of temperatures as judged
by both colors and NaD line strengths. Although a handful of these
stars were previously known, these objects are in general too faint and
too warm to be effectively identified in other modern surveys such as
2MASS, nor are their red/near-IR colors particularly distinctive.  The
implied surface density of FHLCs in this magnitude range is uncertain
at this preliminary stage of the Survey due to completeness
corrections, but is clearly $>$$0.05$~deg$^{-2}$. At the completion of
the Sloan Survey, there will be many hundred homogeneously selected and
observed FHLCs in this sample.

We present proper motion measures for each object, indicating that the
sample is a mixture of extremely distant ($>100$~kpc) halo giant stars, useful
for constraining halo dynamics, plus members of the recently-recognized
exotic class of very nearby dwarf carbon (dC) stars.  The broadband
colors of the two populations are indistinguishable. Motions, and thus
dC classification, are inferred for 40-50\% of the sample, depending
on the level of statistical significance invoked.  The new list of dC
stars presented here, although selected from only a small fraction of
the final SDSS, doubles the number of such objects found by all previous methods. The observed kinematics suggest that the dwarfs
occupy distinct halo and disk populations.

The coolest FHLCs with detectable proper motions in our sample also display
multiple CaH bands in their spectra. It may be that CaH is another
long-sought low-resolution spectroscopic luminosity discriminant between
dC's and distant faint giants, at least for the cooler stars.

\end{abstract}




\section{Introduction}

Although stars with prominent $C_2$ in their spectra have been observed
for more than a century, faint high-latitude carbon stars
(hereafter FHLCs), where here we arbitrarily define ``faint" as $R>13$,
prove to be of very current and special interest for a variety of oddly
unrelated reasons.  Such objects are rare: certainly $<10^{-5}$ of
random stellar images prove to be C stars. Thus FHLCs are, for example,
rarer than QSOs at a given magnitude. They are also not particularly
easy to discover: although the cool N-type stars (with apologies to
\citet{kee93} for the outdated nomenclature) do have very red colors,
the considerably more numerous R and CH stars do not.  Although some
FHLCs are found serendipitously, the majority of past discoveries have
been due to objective prism surveys such as those at Case \citep{san88},
Michigan \citep{mac81}, Kiso \citep{soy99}, Byurakan \citep{gig01},
and Hamburg/ESO \citep{chr01}.
Recent attempts at automated photometric selection of FHLCs have met
with some success for the very red, cool N stars \citep{tot98, tot00,
iba01}, but again the warmer, more numerous FHLCs have still proven
difficult to select autonomously \citep{gre94b}. Although $\sim7000$
galactic carbon stars are known \citep{alk01}, the sum of all the
heterogeneous investigations discussed above has probably yielded
only a few hundred FHLCs.

Why are we interested in finding these rare FHLCs, especially as essentially
all are too faint for the high dispersion spectroscopic analysis that has
been at the core of the study of red giants thus far? It has become clear in
the past decade that the FHLC population consists of two totally distinct,
physically unrelated classes of objects which (confusingly) share remarkably
similar colors and (at least at moderate resolution) spectra. Both of these
two classes are interesting.

Some fraction of the FHLCs are exactly what they appear to be: distant,
luminous evolved giants in the halo.  There have been a small handful of
previous hints that this population extends to rather astonishing distances.
For example, \citet{mar84} serendipitously found one
such star at $d\sim100$ kpc. Clearly the presence of a brief-lived phase of
stellar evolution at these galactocentric distances poses interesting
questions of origin:  could there be star formation in the distant halo, or
in infalling gas? Are these the most luminous members of
previously-disrupted dwarf satellites?  Moreover, aside from the 
question of the origin of
the luminous FHLCs, they make splendid halo velocity tracers \citep{mou85,
bot91}, as at these huge distances they almost surely encompass the entire
dark matter halo, and the very sharp $C_2$ band heads make radial velocity
determinations straightforward even at modest sized telescopes.

The remainder of the FHLCs are perhaps even more exotic. They exhibit large
proper motions \citep{deu94}, and in some cases parallaxes \citep{har98},
that place them at main sequence luminosity ($M_V\sim10$).  These so-called
``dwarf carbon stars", hereafter dC's, should be an oxymoron, as there
should be no way for $C_2$ to reach the photosphere prior to the red giant
phase. For 15 years, precisely one such star was known, G77-61 \citep{dah77,
dea86}, but \citet{gre91} and \citet{gre94a} showed that these are in fact a
surprisingly common subclass of FHLCs, unnoticed in the past simply as most
have $R>16$. A recent review of the dC stars has been given by
\citet{gre00}.  The dozen or so known previous to this work are all at
$d<100$~pc, a volume that contains not a single giant C star.  Therefore,
contrary to the conclusions of 100 years of classical astronomical
spectroscopy, the overwhelming numerical majority of stars with $C_2$ in
their spectra are in fact the previously unknown dwarfs, not giants!
Current thinking is that the $C_2$ in dC's was deposited in a previous
episode of mass-transfer from a now invisible, highly-evolved
companion.  In this respect the dC's are probably similar to barium
stars, and in particular to the so-called ``subgiant CH stars"
\citep{bon74}. Those objects, which despite their names at least occasionally have near main sequence luminosity, are presumably slightly too warm to
show strong $C_2$ despite the inference of $C/O>1$.

Aside from the usual invisibility of the evolved companion, great age for
the dC stars is also implied by the extraordinarily metal poor composition
inferred for the prototype, G77-61 \citep{gas88}. Thus the ultimate
significance of these stars might be to call attention to otherwise
elusive Population~III objects (see also the discussion of \citet{fuj00}).
Furthermore, if an early generation of stars is responsible for
reionization of the intergalactic medium, as now seems increasingly likely
\citep{mad99, fan01, bec01}, then the early Universe achieves a heavy element
abundance already substantially above that inferred for the dC prototype,
perhaps implying that objects such as these are actually pregalactic
\citep{ree98}.

Despite the totally different nature of the giant and dwarf C stars, the
spectra and colors of the disparate classes are frustratingly similar.
Indeed, although some preliminary photometric and spectroscopic luminosity
criteria have been suggested \citep{gre92, joy98}, dC's are currently
identified with complete confidence only if they show detectable 
parallax or proper
motion, thus ruling out membership in the distant halo.

The wide areal coverage ($10^4$~deg$^2$), faint limiting magnitude
($m\sim22$), precision five-color photometry ($\sim0.02$~mag), and highly
multiplexed spectroscopic capabilities of the Sloan Digital Sky Survey
\citep{yor00} provide the opportunity to identify large
numbers of new FHLCs, providing far larger samples than currently
available, both for use as halo dynamic probes and to elucidate the nature
of the engimatic dC's. In addition to merely extending the catalogs of
both classes, major goals of this work are also to develop and/or refine
photometric and spectroscopic luminosity discriminants, hopefully apparent
at low to moderate resolution, and to understand the ratio of dwarfs to
giants in a magnitude limited sample.

Here we report initial results of a search for FHLCs in the SDSS
commissioning data. This subset of SDSS data is very similar, although
not quite identical to, the SDSS Early Data Release (EDR), and the
reader is referred to \citet{sto02} for a detailed description of those
data and their reduction. Our results therefore utilize only $\sim5\%$
of the eventual Survey, but should serve to illustrate the
potential of SDSS to elucidate many issues related to the FHLC problem.
This paper concentrates on methods of selection, and an overview of the
first photometric, spectroscopic, and astrometric results, to give the
reader an understanding of the nature of SDSS FHLC data. A more lengthy
analysis of astrophysical implications will appear in later
publications.

\section{Observations}

\subsection{Selection of Candidates} \label{selection}

Analysis of imaging data from the SDSS camera \citep{gun98, lup01} provides an imaging data base in five broadband filters \citep{fuk96}. The SDSS 
photometric calibrations are described in \citet{hog01} and \citet{smi02}. Unusual stellar objects of a large variety of types are chosen for spectroscopy via automated
analysis of this data base, normally on the basis of odd colors which imply
that the object is interesting.  In the case of FHLCs, the SDSS target
selection algorithm was constructed via observation of a large number of
previously known FHLCs in the SDSS color system \citep{kri98} prior to
the beginning of observations with the SDSS telescope. That work showed
that although R-type FHLCs are normally rather close to the normal stellar
locus in SDSS colors, they may still be separable in a photometric data base that is sufficiently precise and homogeneous. This is not obvious {\it a priori}: the broadband colors of many of these objects are
far from extraordinary, often corresponding to spectral types of mid-K.
\citet{kri98} did not target the very red, cool N stars: although far
more distinct in color space, they are even rarer than the R stars which
are the focus of this paper.

The basic FHLC photometric target selection algorithm currently in use
by the Survey relies mainly on the separation of carbon stars from the
stellar locus evident in the $(g-r)$ vs. $(r-i)$ diagram\footnote{The current, preliminary nature of the SDSS photometric
calibration requires unfortunately cumbersome notation. The SDSS filter
system defined by \citet{fuk96} is denoted $u'g'r'i'z'$, 
but unfortuntaely differs significantly from the filters as realized
on the 2.5m telescope, where the filters and bandpasses are denoted as $ugriz$. However, photometry obtained at this early stage of SDSS is denoted $u^*g^*r^*i^*z^*$ to stress the preliminary nature of
the calibration.} \citep{kri98}.
Objects with stellar morphology, and meeting several criteria that
assure reliable photometry, are selected as candidate carbon stars if they
have $15.0<r^*<19.5$ and fall in the region below $(r^*-i^*) < a + b~(g^*-r^*)$,
where $a=-0.4$ and $b=0.64$.  In addition, candidates targeted for
spectra must be at least as red as $(g^*-r^*)>0.85$, $(r^*-i^*)>0.05$, and
$(i^*-z^*)>0.0$. Candidates with $(g^*-r^*)>1.4$ are given higher priority
for spectra, as they are generally separated further from the stellar
locus in $(g^*-r^*)$ vs. $(r^*-i^*)$. Finally, we also allow for selection
of very red carbon stars ({\it e.g.}, those with dust), considering as
candidates stars with $(g^*-r^*)>1.75$. Note that the precise values of
color-selection regions for carbon stars have varied somewhat during
the SDSS commisioning phase.

The bright cutoff in the selection criteria is imposed by loss of
photometric data due to saturation in the imaging portion of the
Survey, as well as risks of scattered light contamination of fainter
objects during spectroscopy. The most sensitive observations, in $gri$,
saturate at $m\sim14$ in the imaging data. The faint cutoff is 
chosen to yield a spectrum
of reasonable classification quality during the 45-minute Survey
spectroscopic exposures.  This is a fainter limit than
any previous wide-area survey for FHLCs. Even the faintest objects in
our sample are sufficiently well-exposed in the imaging data that there
should be few if any instances of image misclassification ({\it e.g.},
a galaxy selected as a candidate), although we of course have no
defense against visual binaries unresolved in the Survey images.

A further complication, but overall a bonus, in our automated target
selection is that the region in SDSS color space that we select for FHLCs
partially overlaps the target selection region for high redshift
QSOs \citep{ric02}, which have a higher priority for spectroscopic followup in
the Survey. Therefore some candidates which we target as FHLCs have
already been selected for spectroscopy on unrelated grounds. Although
this complicates calculations of the ultimate detection efficiency, it
increases the total yield for at least two reasons: far more spectroscopic
fibers can be allocated to QSO candidates than to FHLCs, and FHLCs which
prove to lie just barely outside of our preset target selection colors
may still lie within the QSO color regions, and therefore be selected
for spectroscopy regardless.
 
Like all other surveys, ours most certainly has selection biases, and
FHLCs of colors markedly divergent from the known ones which shaped
our target selection algorithms may escape detection (although some
may be recovered by the QSO survey, or via ``serendipity"  target selection, which searches
for objects with unusual colors).  Consequently certain derived
quantities from this work, {\it e.g.}, surface densities, must be
regarded as lower limits. However, the results below, which show a
relatively broad range of temperatures in our FHLC sample, as well as
surface densities higher than those derived by previous work, give
us some confidence that our target selection criteria alone are at
least not more biased than previous efforts, and hopefully somewhat
orthogonal thereto. Theoretical evolutionary studies imply a surface
density indicating that previous observations must have missed a
substantial fraction of dC's \citep{dek95}, and it is easy to identify
many potential biases quite aside from limiting magnitude. For example,
previous objective prism selection of FHLCs often required that prominent $C_2$
bands appear on the IIIaJ photographic emulsion, certainly a bias towards
warmer objects. The fact that all three dC's with measured
parallaxes have essentially the identical $M_V, (B-V)$, and $(V-I)$
\citep{har98} is a further hint that current samples are badly biased.

Certainly we expect to recover most or all of the previously known
FHLCs with $m$$>$16 that lie in the SDSS area. Unfortunately these are sufficiently
rare objects previous to our work that there have been only a handful
of such opportunities at this relatively early stage of the SDSS. A further complication is that many objects cataloged in previous surveys are bright enough to cause saturation problems in SDSS. We do recover the N-type FHLC 1249+0146 listed by \citet{tot98}. Another
field already covered in commissioning data that happens to also be
rich with FHLCs is the Draco dwarf spheroidal galaxy. Here we
successfully (and autonomously) recover three previously known FHLCs,
as well as discover a new, previously uncataloged object (see \S \ref{objects}).
However, our target selection algorithms did fail to flag two other
previously known FHLCs that do lie in our data, the relatively bright
($17 < r^* < 18$) stars Draco C2 and C3 \citep{aar82, arm95}. Post-facto
examination of the SDSS photometry for these two stars shows that they
just barely missed selection, lying a few hundredths of a magnitude too
close to the normal stellar locus in $(g-r)$ versus $(r-i)$ space.
The former object {\it was} flagged as a high-z QSO candidate, however,
and may well eventually receive a spectrum on those grounds,
erroneously motivated though they may be, and thus ultimately join the list of successful recoveries. This example illustrates that even
given the precise, homogeneous nature of SDSS photometry, automated
target selection remains an inexact science, and conclusions which
rely on completeness, rather than lower limits on surface density,
must be treated with great caution.

As expected from the target selection criteria (above), we do find that
a major contaminant in our survey for FHLCs (aside from unexpectedly
large photometric errors in otherwise normal stars) are QSOs in the
$2.5 < z < 4$ range \citep{sch02}. A number of DQ white dwarfs 
(degenerate stars with strong $C_2$ bands)
have also been found, and will be discussed elsewhere.

Although the SDSS in spectroscopic mode obtains more than 600 spectra
simultaneously, and $\sim5000$ spectra on one clear winter night, at the end
of the $10^6$ spectra which constitute the project, observations will cease
permanently.  Thus spectroscopic targets are a finite and valuable resource.
However, the rarity of even FHLC candidates, much less actual FHLCs, is such
that this effort is a trivial perturbation upon the SDSS. The photometric
target selection typically yields one candidate FHLC on each 7~deg$^2$
spectroscopic field, and thus the program uses $<0.5\%$ of the spectroscopic
fibers. The completeness of this program is however irrevocably limited by
two further factors, one mechanical and the other programmatic. The minimum
spacing of two spectroscopic fibers projects to $55''$ on the sky, and a
candidate in close proximity to an object with higher scientific priority,
typically a galaxy or QSO candidate, cannot be observed. In addition, there
may be regions where all available fibers for a given spectroscopic exposure
are occupied by higher priority programs.  In both such cases, of course,
promising candidates identified by the target selection algorithms may later
be observed spectroscopically at other telescopes.

\subsection{Spectra of Faint High Latitude Carbon Stars}

At the time of this writing, spectra have been obtained of several
hundred objects that meet the FHLC target selection criteria. A
significant fraction of these objects simultaneously meet the color
selection criteria for other, scientifically-unrelated SDSS programs,
especially high redshift QSOs as noted previously, and were selected
for spectroscopy for those programs rather than specifically to
discover FHLCs.

\subsubsection{Details of Observations}

All but one of the spectra discussed here were obtained with the SDSS
2.5m telescope as part of the spectroscopic survey. These observations,
obtained with a set of plug plates and fiber-fed CCD spectrographs,
cover the 3800 -- 9200 \AA\ range with $\lambda/\Delta\lambda\sim1800$,
and are typically exposed for 45~min (as the sum of three individual
spectra), following constraints set by other SDSS scientific programs.
The reduction and calibration of the spectra are described by
\citet{sto02}.  The objects discussed in this early report were
extracted from the reduced spectra via manual, independent examination
of the spectra by multiple of the authors. In addition, an unpublished
code under development by D. Schlegel selected FHLC candidates via a
comparison with a series of SDSS spectral templates. 
 
A handful of particularly faint photometric candidates were
spectroscopically observed with the Double Imaging Spectrograph of the
ARC 3.5m telescope or the Low Resolution Spectrograph of the
Hobby-Eberly Telescope (HET), although all but one of the final sample
discussed here ultimately received a workable 2.5m spectrum as part of
the continuing SDSS program.

\subsubsection{Identification of New FHLCs} \label{ids}

On the basis of the above spectra, thirty-nine of the SDSS candidates
observed prove to be FHLCs.  These spectra are shown in Figure~1,
and basic astrometric and photometric data are given in Table~1. Most
of the spectra are qualitatively similar to those of the brighter,
objective-prism selected FHLCs in the literature \citep{gre90, bot91, tot98},
dominated by strong $C_2$ Swan bands at $\lambda\lambda4737$, 5165, and
5636, as well as the red CN bands ($\lambda\lambda$7900, 8100). Closer
examination of the spectra, however, shows a richness not present
in previous FHLC investigations: for example, there is a broader range
of NaD strength (presumably due to a larger range of $T_{eff}$) than
present in previous samples, consistent with our freedom from some previous
observational selection effects. In particular, most photographic surveys
in practice required that the Swan bands be well-exposed and prominent
on the $F$ or $J$ emulsion, implying that objects with colors in the
mid-G to mid-K range were normally favored, even if the $C_2$ and CN
bands could be expected over a broader temperature range, as we now
find.

Balmer emission, while not common in C giants,
is certainly not unprecedented, and about the same fraction of our sample
shows H$\alpha$ in emission as the sample of bright giants observed
by \citet{coh79}.  Individually interesting objects will be discussed
in more detail in \S \ref{objects} below.

Figure~2 displays color-color diagrams of the SDSS FHLCs, as well as
that of $\sim10^4$ anonymous faint field stars, to define the normal
stellar locus.  The observed range of $0.8 < (g^* - r^*) < 2.0$ for the
new FHLCs corresponds roughly \citep{fuk96} to $1 < (B-V) < 2$, {\it i.e.}, includes
objects with colors ranging from early K through M stars.  Thus many
of these objects are not extraordinarily red, and are not selected by
methods tuned to red excesses. The FHLC survey of \citet{tot98}, for
example, finds stars chiefly with $(B-V)>2.4$.  The large scatter in the
observed $(u^*-g^*)$ colors in our data is due to the lower sensitivity
of the Survey in the $u$ band  combined with the faintness of most of
the stars in the sample.  Indeed,  many of the stars are not confidently
detected at $u$.  A few objects with unusual colors are noted in \S \ref{objects}.

SDSS coordinates are generally of \decsec{0}{1} rms accuracy in each
coordinate in the ICRS frame, but a large fraction of these stars (see
\S \ref{mu}) prove to have significant proper motions. Therefore we
provide finding charts for our sample in Figure~3.  The epochs of the
images vary slightly, but all lie within one year of 1999.5.

Our automated target selection algorithms have thus far identified few
of the cooler, extremely red FHLCs (presumptively all giants), in
agreement with the expectations from previous work that they are of
relatively low surface density \citep{tot98}. One such example is the
\citet{tot98} object noted in \S \ref{selection}, with $(g^*-r^*)=3.60$
in our data. However, in an unrelated search of very red
objects appearing simultaneously in SDSS and 2MASS, we have found two
further such stars, SDSS~J122740.0-002751 and
J144631.1-005500; they are bright ($r^*=17.03$ and $r^*=15.21$,
respectively), and redder than almost any of our automatically
selected  sample ($(g^*-r^*)=2.76,\ (g^*-r^*)=2.19$). Their spectra,
obtained at the
ESO NTT and displayed in Figure~4, are conventional; the brighter
object displays $H\alpha$ emission, consistent with a giant
classification. As these are not homogeneously selected and may be of
quite different physical nature, we do not discuss these objects
further here (although some basic data appear in Table~2), but simply
note that SDSS is certainly capable of finding these stars; indeed,
their extremely red colors make them easier to select than the more
numerous but anonymously-colored R-type FHLCs that are the focus of
this paper. At the conclusion of the Survey, we may expect a
modest-sized ($\sim10^2$) sample of this type of red object.

\section{Analysis}

\subsection{Radial Velocities} \label{rv}

The precision of the SDSS in derivation of stellar radial velocities
is as yet largely untested. However, two of the stars discussed in this
paper, SDSS~J171942.4+575838 and 172038.8+575934, are recoveries
of previously-known members of the Draco dwarf spheroidal system,
and fortuitously have accurately-measured radial velocities in the
literature \citep{ols95, arm95}, which can then be used to correct the
zero-point of the radial velocities of our new sample.  (A third object in
common described in \S \ref{objects}, Draco C1, is a known radial velocity variable
\citep{ols95} and thus excluded from this calculation.) The resulting
heliocentric velocities
(with 1$\sigma$ uncertainties) are listed in Table~1.  Note that the
velocity difference between these two Draco stars is measured by the above
authors to be $-9.2$ $\rm km~s^{-1}$ and by us as $-12.2$ $\rm km~s^{-1}$,
suggesting that our radial velocity data, or at least this subset thereof,
have quite satisfactory {\it internal} precision. Also, the radial
velocity of the Draco symbiotic SDSS~J171957.7+575005 (see \S \ref{objects}) is given
by the above authors as $-297$ $\rm km~s^{-1}$, while our (corrected)
value is $-312$ $\pm$ 13 $\rm km~s^{-1}$.  We conclude that the radial
velocities shown in Table~1 are accurate to $10-15$~km~s$^{-1}$. We
anticipate later improvements in this precision as the Survey matures.

Even a casual inspection of the radial velocity results suggests that we
are probably observing a mixture of disk and halo populations, as well of
course the already-appreciated mix of stellar luminosities. We use these
preliminary data to derive some inferences on the underlying populations
in \S \ref{pops} below. A preview of one surprising result is appropriate here,
however: the observed magnitude, proper motion (see \S \ref{mu}), and radial
velocity distribution of our sample indicates that, contrary to all
previous FHLC surveys, only a minority of our current stars may be
distant giants.  Therefore at this early stage of the Survey, we make
no attempt to use our radial velocity data for inferences on the
dispersion of the outer halo.

\subsection{Proper Motions and the Dwarf/Giant Ratio} \label{mu}
 
The SDSS Astrometric Pipeline \citep{pie02} computes
J2000 positions for all detected objects, and these positions
can be used for a second epoch for measuring proper motions.
The accuracies for objects with $r^*< 20$ are set by
systematic errors due to short-term atmospheric fluctuations,
sometimes referred to as anomalous refraction, as the telescope
scans along a great circle.  The {\it rms} errors are less than
$\pm$\decsec{0}{1} in each coordinate.
For first-epoch positions, we have taken the position from
the USNO-A2.0 catalog \citep{mon98} for the 21 stars
that are included in USNO-A.  This catalog includes stars
detected on {\it both} the blue and red plates of the first
Palomar Observatory Sky Survey (POSS-I).  The {\it rms} errors are
approximately $\pm$\decsec{0}{17} in each coordinate \citep{deu99},
or somewhat more for faint stars and stars near plate edges.
Because C stars are relatively red, the faint ones are not detected
on the POSS-I blue plates, and are not included in USNO-A.
For these stars, we have used the unpublished positions measured
at the Naval Observatory on all available red survey plates
(POSS-I, SERC, ESO/SRC, and POSS-I reject plates)\footnote{
Many plates were taken for the POSS-I survey and then rejected
from the survey for various reasons.  Some are of poor quality,
but many are of survey quality over most or all of the plate.
These plates have all been measured at the Naval Observatory.}.
The errors depend upon the plate(s) on which a star was detected,
and on the star's brightness.  From the two or more positions and
their epochs, the proper motions and their estimated errors are
calculated and listed in Table~1.  With the typical errors noted
above, and with a typical difference in epoch of 46 years,
the {\it rms} errors are about 6 mas~yr$^{-1}$ in each coordinate
and about 8 mas~yr$^{-1}$ in total motion.

We see from the table that 17 objects in the sample show motions
significant at the $3\sigma$ level, and 26 objects at $2\sigma$
significance. We have considered whether the mere detection of proper
motion requires the inference of dwarf luminosity; one could, for
example, alternatively imagine that both our magnitude and motion limits
are sufficiently sensitive that subgiants might be detected.  For
simplicity we derive crude distances assuming that all dC stars have
$M_{r}$$=$10; our observed motions and radial velocities then yield
total space velocities which are often high (clearly expected due to
selection for detectable proper motion), but {\it all} well below
escape velocity.  Although certainly not definitive, we conclude that
the evidence is consistent with a dwarf classification for each FHLC
with a confidently-detected motion. More quantitative arguments 
in this regard are given in \S \ref{pops}.

If we exclude three objects in the Draco dwarf (see \S \ref{objects}),
the sample numbers 36 objects in total. \citet{gre92} concluded, based on
the first handful of dC stars culled from the heterogenously collected
FHLC lists then available, that $>10\%$ of FHLCs to $V<18$ must be
dwarfs. Given our more homogeneous and, most important, fainter sample,
we now see that this value is probably $\ge50\%$ at $V\le20$. The irony
that this class of dC objects was totally overlooked until quite recently
continues to increase. It has been stated previously by workers in this field
(e.g., \citet{gre00}), but is perhaps still not widely appreciated and thus
bears repeating, that despite the focus of more than one century of astronomical
spectroscopy, the numerical majority of stars with $C_2$ in their spectra
are dwarfs, not giants.

\subsection{Overlap with 2MASS}

At least a handful of very cool, dust-enshrouded carbon stars of giant
luminosity are known to be located in the distant halo; they were discovered
via $JHK$ color selection in ground-based near-IR surveys \citep{cut89,
lie00}. Although it is already clear that SDSS will be complementary to this
past work, and discover predominantly the more numerous, warmer R-type
FHLCs, it is interesting to consider possible overlaps between our FHLC
sample and 2MASS sources \citep{skr97}.  We present results of a search for
our objects in 2MASS in Table~2.

Unfortunately at this very early stage of SDSS, and with the entire
2MASS data set not yet accessible, the number of FHLCs which fall
within areas containing released 2MASS data is still small, and the
list of both detections and non-detections is still clearly limited
by small-number statistics. At the moment one sees merely that, not
surprisingly, the 2MASS detections tend to be the SDSS objects which
are brightest at $z$, and conversely, that SDSS goes so much fainter
than 2MASS for objects with typical FHLC colors that most of our FHLCs will
remain too faint for 2MASS. Of the current positive 2MASS detections,
about half are dC stars (non-zero SDSS proper motions significant at
$3\sigma$), so no obvious luminosity correlation emerges with this very
preliminary sample of cross-identifications.  Although deeper inferences
seem inappropriate until the cross-identified sample (or more probably
upper limits thereon) is much larger, it would seem in accordance with
expectations from previous work that the two surveys are sampling quite
different populations of FHLCs. If one wishes a large sample for halo
kinematic work, the SDSS sample will probably ultimately prevail, as the R
stars are already known to greatly outnumber the N stars at high latitude.
Likewise, as all dC's known to date have R-type spectra, it will almost surely
be SDSS that continues to lengthen the list of carbon dwarfs.

\section{Discussion}

\subsection{Luminosity Indicators in FHLCs: Photometric \& Spectroscopic} \label{indicators}

As noted in the introduction, it is imperative that relatively simple
observational luminosity discriminants be developed for FHLCs. It has
been suspected for a decade that dC stars are substantially more
numerous per unit volume than giants. Our current work now shows
quantitatively that, at least in the $16 < r < 20$ range, a given FHLC
has at least an equal chance of being a dwarf as a giant. As the two
types of star differ in luminosity by $\sim10$~mag, it is particularly
frustrating that discrimination has not proven to be straightforward.

\subsubsection{Photometric Luminosity Discriminants} \label{photo}

Shortly after the realization that dC stars are in fact a common class of
object, \citet{gre92} discussed the possibility that
these objects are often segregated in a $JHK$ color-color diagram from other
stars with C spectra. (Indeed, \citet{dea86} motivated this argument
physically when just one object in the class was known.) Later observational
work by \citet{joy98} and \citet{tot00} provides further encouragement about
the utility of this photometric luminosity indicator. On the other hand,
current data also clearly show that the IR color segregation is imperfect, and
\citet{win96} and \citet{jor98} argue that theoretical model atmospheres for
dC stars imply that, at least if $T_{eff}$ is not independently known, $JHK$
colors alone should not be an unambiguous indicator.

Many or most of the new FHLCs here are bright enough to enable future,
accurate $JHK$ photometry. Even without awaiting further observations,
however, the handful of new dC stars which we report here that fortuitously
lie in the released 2MASS data allows us to make at least a cursory
reassessment of the situation. Table~2 contains four new dC stars
($\mu>0$ with $>3\sigma$ significance) with positive 2MASS detections.
In frustrating conformity with the current ambiguous situation, two of
these objects (SDSS~J073621.3+390725 and 082626.8+470912) exhibit
$(H-K), (J-H)$ colors consistent with the (color-segregated) dC's
discussed by \citet{gre92}, and two (SDSS~J082251.4+461232 and
135333.0$-$004039) are noticeably inconsistent. It seems clear that
$JHK$ photometry is not yet a reliable luminosity discriminant, at
least lacking further ancillary clues.

\subsubsection{Spectroscopic Luminosity Discriminants} \label{spectro}

We have already noted that at low to moderate spectral resolution, giant
and dwarf FHLCs have very similar spectra.  Presumably at high resolution,
there will be various unambiguous discriminants available. However,
these are by definition faint objects, and it would be very useful
to recognize spectroscopic luminosity discriminants accessible to
modest-sized telescopes.  On the basis of the spectra of the first few,
brighter dC's, \citet{gre92} suggested that the appearance of a strong,
sharp $C_2$ bandhead at $\lambda$6191, probably due to the $\Delta v = -2$
bands of $^{13}C^{12}C$ and $^{13}C^{13}C$, might be such a diagnostic
for dwarfs. On the other hand, at least a few contrary examples, 
where AGB stars show the feature prominently, are already known
\citep{gor71, meu01}, so this indicator is clearly not perfectly
reliable.

Our new, spectrally homogeneous sample of FHLCs has been used to reexamine
the issue of the $\lambda$6191 band, and the results are encouraging. Most
of the objects with significant detections of proper motion and good
quality spectra do show the band, sometimes very prominently. For example,
even at the modest scale of Figure~1, the feature is
easily visible in SDSS~J090011.4-003606, and most especially
SDSS~J012150.3+011303 (in fact our highest proper motion object), where it
is very strong. Conversely, most of the good quality spectra of objects
with no confident detection of motion lack the $\lambda$6191 band. However, a
few of the reddest, coolest (as judged by NaD line strength) FHLCs 
with positive motion detections and good quality spectra still lack the $\lambda$6191 band, perhaps implying that the feature may be temperature- as well as luminosity-sensitive. This conclusion, while annoying if correct, 
still detracts little from the
utility of the feature as a luminosity indicator; it merely implies that
lack of $\lambda$6191 in a cooler FHLC is inconclusive regarding the
luminosity. Luckily we are able to suggest below a different indicator for
the cooler subset of FHLCs.

As pointed out in \S \ref{ids}, a small fraction of our objects show
Balmer emission. As dC stars presumably are unlikely to possess active
chromospheres or coronae, or undergo current mass loss, one might look
to this feature as a designator of giant luminosity (especially as the
presumptive binary companion that donated the C is normally invisible
in the spectrum).  Unfortunately, we already know of contrary cases:
the dC PG0824+289 shows Balmer emission \citep{heb93}, presumably due
to heating of the dwarf by the very close hot DA companion which is
also visible in the spectrum. We draw attention to a second
possible case in this class in \S \ref{objects} below.

Several of our better exposed spectra show prominent BaII
$\lambda\lambda$6130, 6497 absorption (the latter normally blended with
Ti, Fe, and Ca); an example is given in Figure~5.  Although normally
prominent in supergiants, Ba is known to be enhanced in dC stars as
well \citep{gre94a}. Therefore although this is an astrophysically
interesting feature, it probably will not be a reliable luminosity
indicator.

An interesting feature of some of our spectra is the presence of very
prominent CaH $\lambda\lambda$6382, 6750 bands in the cooler FHLCs,
i.e., those with strong NaD and redder colors. Again a good example is
shown in Figure~5. Although we are not aware of any previous discussion
of this feature in C stars, in K and M stars CaH is normally strong
only in dwarfs, peaking at late K (e.g., \citet{kir91}). We propose
that for this cooler subset of FHLCs, CaH may be an effective low
resolution luminosity indicator -- in our sample, only stars with
positive motion detections show the bands. Serendipitously, this
feature appears at temperatures which are apparently too cool for the
$\lambda$6191 $^{13}C^{12}C$, $^{13}C^{13}C$ band discussed above, and
is thus nicely complementary.

Despite the size of our current sample as compared with previous work,
when subdivided by temperature and proper motion, the number of
FHLCs with high quality spectra is still quite modest. We defer
until a subsequent paper, where it is already evident that the sample
size will more than double, more quantitative discussions of
the luminosity and temperature correlation of spectral features.
Ultimately photometric variability may prove a simple luminosity
discriminant, although the absence of same will remain inconclusive. While
a few isolated measures which show variability could still be due to
interactions in the rare very close binary dC's such as PG0824+289,
consistent, chaotic variability is probably associated only with mass
loss in the giant FHLCs.

\subsection{Surface Density of FHLCs}

As discussed briefly in \S \ref{selection}, our current sample of 39 carbon stars 
should not be considered ``complete" for several reasons. Most importantly, 
SDSS galaxies and quasar candidates are targeted for follow-up 2.5m 
spectroscopy at higher priority than the bulk of the carbon star candidates. 
In some regions of the survey, the surface density of galaxies and QSO 
candidates is so high as to effectively consume nearly all available 
spectroscopic fibers. On average, in the early commissioning phases of the 
survey, about 40\% of the requested carbon star candidates received 
spectroscopic fibers, but even this average number changed as spectroscopic 
target selection algorithms for various other object classes were refined during 
SDSS commissioning. On the other hand, some of the carbon stars discussed 
in this paper are fainter than the typical $r<19.5$ limit imposed 
for 2.5m spectroscopy for the bulk of SDSS carbon star candidates. Aside from 
their magnitudes, these fainter discoveries do generally meet the same 
color-selection criteria for other carbon star candidates (as may be discerned 
from Figure 2), but in fact were selected for spectra by the algorithms aimed 
at high redshift quasar candidates (for which the magnitude limit is somewhat 
fainter). 

We may therefore only quote a conservative lower limit on the number
density of confirmed carbon stars (with $r>15$; the SDSS spectroscopic
bright limit), from the following considerations. The region surveyed
spectroscopically with the SDSS 2.5m, at the time of compilation of the
current sample, is of order 700 deg$^2$. In that region, SDSS found at
least 35 confirmed carbon stars; we conservatively exclude from this
count the 3 carbon stars in Draco and the 1 object spectroscopically
confirmed from the ARC 3.5m. Thus the surface density of FHLCs is
$>$$0.05$~deg$^{-2}$. This number differs little from the value reported
by the recent Hamburg/ESO Survey for FHLCs \citep{chr01}, despite the
fact that SDSS clearly can reach objects 3~mag fainter. 
This is a striking and surprising result, but
it is much
too early in our survey to attempt to attach any physical significance
to this issue: incompleteness of SDSS selection is at least
as likely an explanation as any genuine flattening in the faint end of the
luminosity function. Regardless of these current ambiguities, it
is clear that SDSS will discover and spectroscopically
confirm several hundred, and perhaps a few thousand, FHLCs 
in the full survey. Far more quantitative limits on FHLC surface density
will therefore become available when a substantial volume of SDSS production
data are on hand. At the time of submission of this paper, an additional
100 FHLCs have been identified, but not yet examined in detail.

\subsection{Some Comments on Population Issues} \label{pops}

The FHLCs selected from the APM survey \citep{tot98, iba01}
are cool AGB stars at distances of tens of kpc.
The stars found in this work, however, are primarily
warmer and fainter.  Both differences could result in different
mixes of giant and dwarf stars, or of halo and disk stars,
in our sample.  In order to place some constraints on the
population of stars in our sample, we have constructed 
simple Monte-Carlo models of possible FHLC populations, to ascertain
which models are consistent with the observed sample and which
are not. These models are not exhaustive and will be developed
more thoroughly in the future as the sample size increases.

\subsubsection{Models}

The Monte Carlo program places stars around the Sun based
on an assumed space density (with an exponential scale height
for disk stars and a galactocentric radial power law for halo stars),
with absolute magnitudes drawn from an input luminosity function (LF),
and with UVW space velocity components drawn from an input
distribution of a gaussian velocity ellipsoid with specified UVW
dispersions and Galactic rotation velocity.
Then, for stars not in the Galactic plane with apparent magnitudes
within a specified range ($15.0 < r^* < 19.5$), the observables
(proper motion, radial velocity, and apparent magnitude)
including observational errors appropriate for SDSS data
are calculated and are entered in the model sample.

The program requires values for the velocity ellipsoid and the LF.  The
input LF used for giant C stars has a maximum per unit magnitude at
$-3.5 < M_r < -2.5$, with progressively fewer stars at $-4.5 < M_r <
-1.5$.  This LF is fainter than the typical $M_R \sim -3.5$ expected
for stars in the APM survey \citep{tot00}, but is consistent with the
warmer colors of stars in our sample (see their Figure~4).  The LF used
for all dwarfs has a maximum at $9 < M_r < 10$, with progressively
fewer stars at $7 < M_r < 12$.  This is consistent with the three dwarf
C stars with measured parallaxes, all of which have $M_r \sim 9.3$
\citep{har98}.  Values for the velocity ellipsoid are taken from
\citet{rei95} for the disk, and from \citet{chi00} for the halo and
thick disk.  A scale height of 250~pc is used for the disk, and 1000~pc
for the thick disk.

\subsubsection{Results of Modeling}

The models make predictions for the distributions of apparent
magnitude, radial velocity, and proper motion that can be compared with
the observed distributions. The three known extragalactic stars
(members of the Draco dwarf) are excluded from the comparison, as well
as two other objects where radial velocity data were not available at
the time of the analysis, or where uncertainties were abnormally large
due to reduction problems, leaving the final comparison sample at 34
objects.

The results are shown in Figure 6, where the top figures show several
illustrative models, and the bottom figures show a model with three
components that is consistent with all the data.  The apparent magnitudes
are shown in panel (a).  The falling curve predicted for halo giant stars
is due to the space density of halo stars dropping steeply in the outer
halo.  (A density $\propto r^{-3.5}$ is often accepted for the Galaxy.)
Panel (a) indicates that most stars in the sample are probably dwarfs,
and that the fraction of dwarfs is likely to be rising toward fainter
magnitudes.  This result is completely opposite from the conclusion
of \citet{tot00} on the dwarf/giant ratio in the APM survey of FHLCs.
The APM survey, however, is made up of much redder and (especially)
brighter stars with a magnitude distribution dropping toward fainter
magnitudes -- it has only two objects with $R > 16$, whereas our sample
has essentially all stars of $r^* > 16$ -- so the markedly different conclusions
are perhaps not unexpected.

The distribution of radial velocities is shown in panel (b).
The middle panels indicate that about two thirds of the sample
is a halo population, either dwarf or giant stars.
This substantial fraction of disk dwarf stars is larger than for the
halo-dominated dwarf carbon stars already known, but the difference
is not unexpected given that the selection is not kinematically
biased.

The distribution of proper motions depends on the distances
of the stars, and their inferred distances depend on the assumed luminosity.
Because the luminosity function of dwarf carbon stars
is unknown, different proper motion distributions can be fit
by adjusting the assumed LF.  Choosing the LF to fit the data is,
in essence, finding the statistical parallax that makes the radial
velocity and proper motion distributions mutually consistent.
Panel (c) shows results, using a single LF (see below) for all dwarfs.
It is possible that disk and halo dwarf carbon stars have different
LFs.  The fit to the observed distribution would be improved if
the disk dwarfs have lower luminosities than the halo dwarfs.
The LF for giant stars is better known, and it has little impact
because the proper motions are small in any case.

The lower panels compare our sample data (excluding the three stars
in Draco that are not representative of the Galactic halo) with
a simple population model with three components that fits the data
fairly well: 15\% halo giants, 50\% halo dwarfs, and 35\% disk dwarfs.
Other combinations are also consistent with the data within limits;
e.g. disk stars can be replaced with thick disk stars, and the proportion of
halo and disk dwarfs and their luminosities can be changed in tandem. The
conclusion is inescapable, however, that the dC stars are commonly found in
both halo and disk populations.  This conclusion has been anticipated on
general stellar evolution grounds \citep{dek95}, but the previous small,
heterogeneous samples of dC's have yielded scant empirical evidence
in this regard.

For the stars in our sample with statistically significant proper
motions, the reduced proper motions are in the $14 < H_r < 19$ range,
and upper limits for the stars without significant motions range from
$H_r < 13$ through $H_r < 16$.  These results are also consistent with
a mixture of halo dwarfs, disk dwarfs, and giants, but a larger sample
will be needed to make more cogent comments.

Our models imply that whatever giant stars exist in our sample must
(not surprisingly) have insignificant proper motion, and most should be
bright with large radial velocity.  (Note that the three giants in
Draco do match these expectations.)  The best candidate on these
simplisitic grounds is SDSS~J114125.8+010504, whose blue colors and
spectrum do in fact strongly imply it is a giant (see \S
\ref{objects}).  Other less likely candidates are
SDSS~J095516.4+012130, J221854.3+010026,  J144150.9-002424,
J075116.4+391201, and  J015232.3-004933.  Indeed, probably only a
handful of the current sample are giants.  For some time it has been
realized that there must be some threshold sensitivity where dC stars
become more the norm rather than a rare curiosity amongst carbon
stars:  it appears that the SDSS has crossed this threshold.

One concern about this discussion is that the sample may be biased
toward one population or another by selection effects that are not
included in the models.  In this initial paper, some biases undoubtedly
exist toward some spectral types, temperatures, and atmospheric
compositions that we do not yet understand.  We know of no reason why
selection of dwarfs should be favored over giants, for example.  It is
plausible, however, that warm metal-poor dwarfs may be favored over
warm metal-rich dwarfs, creating a higher halo/disk ratio in the sample
than really exists in the solar neighborhood.  Future work with a
larger sample should help to address these issues.

\section{Comments on Individual Objects} \label{objects}

SDSS~J003813.2+134551. $H\alpha$ in emission.

SDSS~J012150.3+011303. This star has by a considerable margin the largest
proper motion in our sample (\decsec{0}{24}~yr$^{-1}$), and amongst the
largest of any dC yet reported (cf. \citet{deu94}). At this magnitude
($r^*=17.0$) in the northern sky, it would be surprising if it had not
already been noted on the grounds of motion alone.  This proves indeed to be
the case; it is one of the few previously-cataloged objects in our sample.
The object is identical to LP587-45, but has a confusing and unfortunately
erroneous literature trail which we attempt to unsnarl here.  LP587-45 was
proper motion selected by \citet{luy79a} in the NLTT catalog, where
photographic magnitudes and proper motion data are in excellent agreement
with our results. (The precise agreement of the motion and position angle
make the cross-identification unambiguous.) The NLTT lists the color class
as ``$m$" (essentially simply ``red"), and to our knowledge there has been
no previously reported spectrum, and thus recognition of the unusual dC
status.  Unfortunately, \citet{luy79b} designates this star as the visual
companion of the far-more famous LP587-44 \citep{luy80} (=~WD0119-004, =
GR516), a very-well studied bright DB star with numerous literature
citations. This attribution is in error; LP587-45 and LP587-44 have
essentially the identical right ascensions, but differ in declination by
$\sim 1.5^{\circ}$.  Inspection of the POSS indicates that the DB star does
indeed have a faint companion of $9''$ separation at $p.a.=264^{\circ}$ as
recorded by \citet{luy80}, but this companion should be designated as
anonymous, or LP587-44B, not the (now recognized) dC star LP587-45, an
entirely unrelated object in a different part of the sky.

SDSS~J013007.1+002635. One of the few objects with strong, narrow $H\alpha$
emission, and one of the coolest in the sample (very strong NaD); more
important, this star has a composite spectrum. The object is probably a
dwarf, with a $2.3\sigma$ detection of proper motion. In addition to the
Swan and red CN bands, there are multiple broad Balmer absorptions in the
blue. We are aware of only two previously-reported dC stars with composite
spectra \citep{heb93,lie94}. However, the situation here is complex; the SDSS images
show a fainter, blue companion located \decsec{4}{5} WNW. It is possible
this object contaminates the spectrum of the FHLC. On the other hand, the
photometry from the imaging data base, where the two objects should be
well-separated, shows the FHLC to be far more ultraviolet than any other
object in the sample, perhaps suggesting that the Balmer and $C_2$ features
do indeed originate from the same image.  Further observations are clearly
needed.

SDSS~J033704.0-001603. Strong $H\alpha, H\beta$ emission.

SDSS~J085853.3+012243. Contradictory diagnostics: strong $H\alpha,
H\beta, H\gamma$ emission; very prominent NaD, CaH absorption. Significant
proper motion detection and thus a rare dC with emission, perhaps
indicating an unseen but hot, irradiating companion. Yet the observed
$(g^*-r^*)=2$ is the reddest object in the sample.

SDSS~J112801.7+004035. Note extremely high radial velocity; surely a
candidate extragalactic object, e.g., a member or ex-member of a dwarf spheroidal in the Local Group. Colors and spectrum are normal.

SDSS~J114125.8+010504. Stands out as the bluest object in the sample in
$(g^*-r^*)$, and very red in $(i^*-z^*)$. Spectrum of a classical CH
star, similar to HD~5223 (see \citet{bar96}).  High radial velocity and
lack of proper motion perhaps suggest halo giant system.

SDSS~J132840.7+002717. Noisy spectrum: classification may be uncertain.

SDSS~J171942.4+575838. Previously known C star in the Draco dwarf galaxy
(=BASV~461, =Draco~461); see \citet{baa61} and \citet{arm95}.

SDSS~J171957.7+575005. Previously known, highly unusual emission
line object in the Draco dwarf galaxy (= Draco~C1). First reported
by \citet{aar82}, further remarks by \citet{mun91}, \citet{arm95} and
\citet{ols95}. It is generally classified as a symbiotic carbon star,
of which fewer than a handful are known both in the Galaxy and Local Group
combined. Although the prototype, UV~Aur, has been studied for some time
\citep{san49}, objects like this one with excitation as high as indicated
by the very strong HeII $\lambda$4686 emission are almost unknown; the
other examples such as SS~38 \citep{sch88} and UKS~Ce-1 \citep{lon77}
are hardly household words. Although previously detected in X-rays
\citep{bic96, mur96}, our automated pipeline processing system also
drew attention to the coincidence with a ROSAT X-ray source, indicating
that new stellar X-ray source optical identifications will be made
by SDSS.  Our spectrum is of greater resolution, wavelength coverage,
and signal-to-noise ratio than previously published spectra; in an effort to
stimulate further interest in this extraordinary star, we display the
spectrum of Figure~1 at a greatly enlarged scale in Figure~7, truncating
the strong H$\alpha$ emission so that other features are readily visible.
Excess blue flux in our photometry, possibly due to the continuum of
the companion.

SDSS~J172038.8+575934. Previously known C star in the Draco dwarf
galaxy (=BASV~578, =Draco~C4); see \citet{baa61}, \citet{arm95} and
\citet{ols95}.

SDSS~J172909.1+594035. Extreme colors: the bluest $(u^*-g^*)$ and
$(i^*-z^*)$ of the sample (see Table~1, Figure~2), but  very faint, and
the uncertainties on the measured magnitudes are large. Although located
$\sim2.4^\circ$ from the center of the Draco dwarf, whose tidal radius is
only $\sim0.65^\circ$ \citep{ode01}, it has the correct radial velocity
for membership.  If a member, the implied luminosity is low compared
with the other known Draco C giants, but the object could be a CH star,
and the (quite noisy) spectrum does not contradict this possibility.
The issue of possible tidal debris far from Draco has recently been
considered by \citet{pia01} and \citet{ode01}, who find no candidate
(ex-)members nearly this distant, although they search for macroscopic
overdensities and color sequences, rather than individual stars; and
those Draco candidate members that are found relatively distant from the core
are near the major axis, contrary to this object. Certainly other Local
Group dwarfs such as the Sagittarius dSph are known to have debris many
degrees distant (\citet{new02} and references therein), so an asssociation here may still be plausible and
important.
 
\section{Conclusions}

Despite their rarity and the relatively benign colors of the majority
of the objects, large numbers of FHLCs can be efficiently selected by 
the SDSS. The current sample, although small compared with
the ultimate end product of the Survey, already provides interesting
information on a variety of FHLC issues. At the completion of the
Survey, the homogeneously selected FHLC sample will for the first time
be sufficiently large that statistics are no longer dominated simply by
the size of the catalog, though a negligible fraction of SDSS resources
are applied to this problem.

The SDSS has already fulfilled the previously undemonstrated expectations
that a sufficiently sensitive and efficient selection technique for
FHLCs would yield the (formerly) exotic dwarf carbon stars in copious
numbers. For the first time we constrain the ratio of carbon dwarfs
to giants in the $16 < r^* < 20$ range: it is at least near to unity,
and possibly considerably larger. Preliminary kinematic analyses imply
that there are distinct halo and disk dwarf populations.

Lacking the positive detection of proper motion, separation of the
dwarfs from giants for FHLCs remains problematic. We find no single
photometric or (low resolution) spectroscopic diagnostic that applies
to all objects, but suggest the addition of one further weapon to the
arsenal: in sufficiently cool FHLCs, the presence of strong CaH bands
is an effective dwarf indicator.

Numerous unusual FHLCs have also been identified for further study.

\acknowledgments

We are indebted to Jim Liebert and especially Paul Green
for countless useful discussions about FHLCs. We thank J.~A.~Smith,
J.~MacConnell, G.~Wallerstein, and H.~Bond for helpful comments.
One of us (B.M.)  thanks the NASA Goddard Space Flight Center for
hospitality during some of this work. Portions of this paper are based
on observations obtained with the Apache Point Observatory 3.5-meter
telescope, which is owned and operated by the Astrophysical Research
Consortium, on observations collected at the European Southern Observatory, Chile, for proposal \#67.A-0544, and data from 
the Hobby-Eberly Telescope (HET), a joint project of
the University of Texas at Austin, the Pennsylvania State University,
Stanford University, Ludwig-Maximillians-Universit\"at M\"unchen, and
Georg-August-Universit\"at G\"ottingen.  The HET is named in honor of
its principal benefactors, William P. Hobby and Robert E. Eberly.
This publication makes use of data products from the Two Micron All
Sky Survey, which is a joint project of the University of Massachusetts and 
the Infrared Processing and Analysis Center/California Institute of
Technology, funded by NASA and the NSF. This research has made use of 
the SIMBAD database, operated at CDS, Strasbourg, France. 
IRAF is distributed by the National Optical Astronomy Observatories,
which are operated by the Association of Universities for Research
in Astronomy, Inc., under cooperative agreement with the National
Science Foundation. We acknowledge the support of NASA Grant 
NAG5-11094 (G.~K.), NSF Grants AST-0071091 (M.~A.~S.), 
AST-9900703 (D.~P.~S.), and PHY-0070928
(X.~F.), Chilean grant FONDECYT/1010981 (P.~B.~H.), 
as well as a Frank and Peggy Taplin Fellowship (X.~F.).

The Sloan Digital Sky Survey (SDSS) is a joint project of The University
of Chicago, Fermilab, the Institute for Advanced Study, the Japan
Participation Group, The Johns Hopkins University, Los Alamos National
Laboratory, the
Max-Planck-Institute for Astronomy (MPIA), the Max-Planck-Institute for
Astrophysics (MPA), New Mexico State University, Princeton University, the
United States Naval Observatory, and the University of Washington. Apache
Point Observatory, site of the SDSS telescopes, is operated by the
Astrophysical Research Consortium (ARC).  Funding for the project has been
provided by the Alfred P. Sloan Foundation, the SDSS member institutions,
the National Aeronautics and Space Administration, the National Science
Foundation, the U.S. Department of Energy, the Japanese Monbukagakusho, and the Max Planck Society. The SDSS Web site is http://www.sdss.org/.

%% Use LaTeX's the bibliography environment to mark up your reference list.



%% Generally speaking, only the figure captions, and not the figures
%% themselves, are included in electronic manuscript submissions.
%% Use \figcaption to format your figure captions. They should begin on a
%% new page.

\clearpage

\oddsidemargin = -1cm

%\begin{footnotesize}
\begin{deluxetable}{lcccccccc}
\tabletypesize{\scriptsize}
\tablenum{1}
\tablecolumns{9}
\tablewidth{0pc}
\tablecaption{Faint High Latitude Carbon Stars Discovered in SDSS Commissioning Data}
\tablehead{
Object Name$^{a}$ & $r^*$ & $(u^*-g^*)$ & $(g^*-r^*)$ & $(r^*-i^*)$ & $(i^*-z^*)$ & $\mu$ 
& p.a. & RV \\
SDSS~J+ & & & & & & (mas~yr$^{-1}$) & $^\circ$ & (km~s$^{-1}$) 
}
\startdata
    003013.1$-$003227 & 19.49 & 2.84 & 1.87 & 0.50 & 0.24  & 11(9)  & ...  & 46.3(8.3) \\
    003813.2+134551 & 19.32 & 2.22  & 1.60 & 0.50 & 0.31 & 5(12)  & ...  & $-10.0$(11.3) \\
    003937.3+152911  & 18.61 & 1.67 & 1.05 & 0.30 & 0.17 & 31(8)  &  174 & $-130.1$(17.1) \\
    012150.3+011303  & 17.02 & 2.71 & 1.69 & 0.50 & 0.14 & 241(8) &  124 & $-18.3$(7.7) \\ 
    012526.7+000449  & 19.34 & 1.42 & 1.63 & 0.54 & 0.16 & 56(12) &  101 & $-49.1$(17.3) \\
    013007.1+002635  & 17.62 & 0.40 & 1.01 & 0.43 & 0.36 & 28(12) &  135 & $-16.2$(9.4) \\                 
    015232.3$-$004933  & 18.28 & 3.22 & 1.56 & 0.57 & 0.29 & 19(8)  &  ... & $-31.2$(11.3) \\
    023208.6+003639  & 19.46 & 2.00 & 1.59 & 0.56 & 0.27 & 34(12) &  130 & 9.6 (21.2) \\
    025634.6$-$084854  & 19.47 & 1.94 & 1.10 & 0.25 & 0.17 & 43(12) &  143 & $-121.0$(12.0) \\
    033704.0$-$001603  & 18.67 & 2.38 & 1.92 & 0.71 & 0.30 & 11(8)  &  ... & $-25.4$(12.0)$^b$ \\                 
    073621.3+390725  & 18.43 & 3.21 & 1.52 & 0.49 & 0.16 & 58(8)  &  234 & $-28.7$(11.3) \\
    075116.4+391201  & 17.61 & 3.47 & 1.60 & 0.58 & 0.43 &  7(8)  &  ... & $-15.3$(9.9) \\
    075953.6+434021  & 19.48 & 3.64 & 1.91 & 0.63 & 0.13 & 6(10)  &  ... & 6.3(26.6) \\
    082251.4+461232  & 17.22 & 2.79 & 1.40 & 0.77 & 0.22 & 25(8)  &  219 & 47.7(8.0) \\
    082626.8+470912  & 17.77 & 2.80 & 1.44 & 0.48 & 0.31 & 41(8)  &  143 & 49.4(6.6) \\
    085853.3+012243  & 18.31 & 2.46 & 1.99 & 0.67 & 0.36 & 29(8)  &  130 & 89.8(16.2) \\
    090011.4$-$003606  & 18.44 & 3.29 & 1.47 & 0.47 & 0.26 & 59(8)  &   3  & 227.1(12.5) \\
    094858.7+583020  & 18.77 & 2.55 & 1.15 & 0.33 & 0.22 & 24(8)  &  176 & 
$-29.8$(42.0) \\
    095516.4+012130  & 18.35 & 2.85 & 1.60 & 0.52 & 0.26 & 19(8)  &  ... & 154.3(15.4) \\    
    100432.5+004338  & 19.99 & 1.64 & 1.71 & 0.56 & 0.19 & 36(12) &  174 & 167.4(17.9) \\                
    112801.7+004035  & 18.81 & 2.44 & 1.41 & 0.38 & 0.23 & 22(10) &  ... & 562.1(23.3) \\ 
    112950.4+003345  & 18.28 & 2.26 & 1.40 & 0.41 & 0.33 & 44(10) &  236 & 35.8(10.8) \\
    114125.8+010504  & 17.29 & 2.12 & 0.89 & 0.15 & 0.29 & 4(8)   &  ... & 119.8(10.8) \\
    114731.7+003724  & 18.60 & 1.70 & 1.05 & 0.23 & 0.13 & 34(8)  &  179 & 43.3(5.2) \\
    115925.7$-$031452  & 18.76 & 1.78 & 1.08 & 0.17 & 0.10 & 23(10) &  ... &   ...$^c$ \\
    132840.7+002717  & 19.40 & 1.88 & 1.17 & 0.18 & 0.18 & 47(16) &  165 & 117.7(11.0) \\
    135333.0$-$004039  & 16.60 & 2.68 & 1.59 & 0.53 & 0.27 & 65(8)  &  235 & $-32.1$(13.8) \\
    142112.4$-$004823  & 19.05 & 2.13 & 1.51 & 0.36 & 0.42 & 23(7)  &  105 & $-85.0$(9.0) \\
    144150.9$-$002424  & 17.86 & 2.56 & 1.72 & 0.66 & 0.26 & 8(8)   &  ... & $-44.8$(14.4) \\
    153732.2+004343  & 17.63 & 2.54 & 1.82 & 0.63 & 0.38 & 36(8)  &  155 & 56.4(8.4) \\
    161657.6$-$010350  & 19.06 & 2.70 & 1.70 & 0.60 & 0.22 & 25(10) &  215 & $-46.3$(11.4) \\                      
    171942.4+575838  & 16.81 & 3.29 & 1.18 & 0.33 & 0.34 & 19(8)  &  ... & $-301.4$(13.1) \\   
    171957.7+575005  & 16.47 & 1.25 & 1.25 & 0.28 & 0.41 & 5(8)   &  ... & $-311.7$(13.5) \\    
    172038.8+575934  & 17.76 & 2.65 & 1.17 & 0.29 & 0.23 & 6(8)   &  ... & $-289.2$(10.3) \\  
    172909.1+594035  & 20.19 & 0.12 & 1.84 & 0.44 &$-0.09$ & ...$^d$ &  ... & $-281.9$(16.7) \\  
    173650.5+563801  & 19.19 & 2.27 & 1.72 & 0.67 & 0.32 & 18(12) &  ... &  $-19.2$(10.5) \\  
    221450.9+011250  & 19.85 & 1.86 & 1.54 & 0.55 & 0.11 & 34(15) &  219 &  $-191.7$(19.0) \\
    221854.3+010026  & 19.24 & 1.78 & 1.09 & 0.37 & 0.15 & 18(12) &  ... &  $-307.2$(10.2) \\   
    230255.0+005904  & 17.71 & 2.30 & 1.45 & 0.52 & 0.31 & 46(8)  &  236 & $-34.6$(13.1) 
\enddata 
\tablenotetext{~} {General notes: Parenthesized values are $\pm1\sigma$ uncertainties. See \S \ref{objects} for specific notes on individual objects.
Radial velocities are heliocentric.}
\tablenotetext{a} {Naming convention is equinox 2000, $hhmmss.s,
\pm ddmmss$. Epochs vary slightly, but all lie in the $1999.5\pm1.0$ interval.}
\tablenotetext{b} {Two separate observations exist; weighted average RV is quoted.}
\tablenotetext{c} {ARC 3.5m spectrum only; no RV data available.} 
\tablenotetext{d} {Object very faint; no suitable first epoch data for quantitative proper motion analysis, although informal indications are that any motion must be small.}
\end{deluxetable}
%\end{footnotesize}

\clearpage

\oddsidemargin = 0cm

%\begin{footnotesize}
\begin{deluxetable}{lccccc}
\tabletypesize{\scriptsize}
\tablenum{2}
\tablecolumns{6}
\tablewidth{0pc}
\tablecaption{SDSS Faint High Latitude Carbon Stars in Released 2MASS Fields}
\tablehead{SDSS~J+ & 2MASS+$^a$ & $z^*$ & J & H & K } 
\startdata
   003013.1$-$003227   & ... & 18.75 & ... & ... & ...   \\
   003813.2+134551     & ... & 18.51 & ... & ... & ...   \\   
   003937.3+152911     & ... & 18.14 &... & ... & ...   \\   
   012526.7+000449     & ... & 18.64 & ... & ... & ...   \\  
   025634.6$-$084854   & ... & 19.03 & ... & ... & ...   \\    
   033704.0$-$001603   & 0337040$-$001602 & 17.66 & 16.25(0.09) & 15.47(0.09) & 15.15(0.14) \\
   073621.3+390725     & 0736213+390725 & 18.10 & 16.55(0.13) & 16.04(0.20) & 15.60(0.22) \\
   075116.4+391201     & 0751163+391201 & 17.46 & 15.18(0.06) & 14.40(0.06) & 14.10(0.07) \\
   075953.6+434021     & ... & 19.03 & ... & ... & ...   \\ 
   082251.4+461232     & 0822514+461231 & 16.68 & 15.50(0.07) & 14.6 (0.08)  & 14.49(0.07) \\
   082626.8+470912     & 0828267+470911 & 17.59 & 15.96(0.10) & 15.22(0.11)  & $>$14.61  \\   
   090011.4$-$003606   & ...$^b$  & 17.71 & ... & ... & ... \\           
   115925.7$-$031452   & ... & 18.49 & ... & ... & ...   \\    
   122740.0$-$002751$^c$ & 1227400$-$002751 & 15.08 & 12.76(0.03) & 11.50(0.02) & 10.55(0.03) \\
   135333.0$-$004039   & 1353330$-$004039 & 16.34 & 14.59(0.04) & 13.80(0.04)  & 13.61(0.05) \\
   144150.9$-$002424   & 1441509$-$002424 & 17.46 & 15.50(0.07) & 14.67(0.07)  & 14.22(0.08) \\
   144631.1$-$005500$^c$ & 1446311$-$005500 & 14.96 & 12.38(0.03) & 11.53(0.03) & 11.05(0.03)
\enddata 
\tablenotetext{~} {General notes: Parenthesized values are $\pm1\sigma$
uncertainties. $z^*$ magnitudes from SDSS; $JHK$ magnitudes from 2MASS Second Incremental Release Point Source \hbox{Catalog.} Note that the former system is $AB$-based, and the latter, Vega-based.}
\tablenotetext{a} {Blank entry indicates object lies within 2MASS survey area, but was not detected by 2MASS.}
\tablenotetext{b} {Faintly visible on sky image gifs, but not cataloged by 2MASS.}
\tablenotetext{c} {Not part of autonomously selected sample, but included here for completeness: see \S \ref{ids}.}
\end{deluxetable}
%\end{footnotesize}

\clearpage

\figcaption{Spectra of 39 high-latitude carbon stars observed by SDSS.
With the exception of 115925.7-031452, all data were obtained with the
SDSS 2.5 m telescope and spectrographs; the latter was observed by the
APO 3.5 m telescope. The spectra are sky-subtracted and flux-calibrated
using observations of standard early F stars.  The results of imperfect
sky subtraction are seen at 5577 \AA{} and at wavelengths $>8500$ \AA{}
in many of the spectra.  The data are binned to a resolution of about
5 \AA{} for the brighter stars and 10--20 \AA{} for the fainter objects.
\label{fig1}}

\figcaption{Color-color diagrams of the 39 SDSS FHLCs, as well as
SDSS photometry of $\sim14,0000$ anonymous field stars with $r^*<17$
\citep{fin00}, displayed to illustrate the normal stellar locus. The
new objects are displaced from the normal locus in the direction and
amount predicted by \citet{kri98} on the basis of observations of
previously known FHLCs, but the relatively small degree of segregation
demonstrates that the photometry must be both precise and homogeneous
to discover FHLCs with high efficiency. {\it Open symbols:} carbon
stars with $\mu>30$~mas~yr$^{-1}$, and thus newly recognized dwarfs;
{\it filled symbols:} stars with proper motions in SDSS data of
$\mu<30$~mas~yr$^{-1}$, and thus probably a mixture of giants and dwarfs.
Clearly SDSS colors do not provide effective luminosity discriminants. A
few objects with unusual colors are discussed in \S \ref{objects}.
\label{fig2}}

\figcaption{Finding charts for the~39 faint high latitude carbon stars
observed in SDSS commissioning data.  All frames are~$i$-band images taken
with the SDSS camera, and are~100$''$ on a side.  The small arrow in
the lower left of each chart indicates the direction of north; all
charts have ``sky" parity, so east is located~90$^{\circ}$
counterclockwise from the north arrow.  \label{fig3}}

\figcaption{ESO NTT spectra of two very red FHLCs found in SDSS data. These
spectra, obtained in April 2001, are of spectral resolution $R\sim270$
({\it lower panel}) and $R\sim600$ ({\it upper panel}). These are probably
the rarer, N-type (giant) FHLCs discussed by \citet{tot98} rather than
the R-type objects copiously found in our survey; note the $H\alpha$
emission in the brighter object. Unlike SDSS spectra, the telluric bands
are not removed in these reductions; the A and B bands are particularly
prominent. The absolute flux calibration is uncertain outside of the
$\lambda\lambda5300-9300$ range.
\label{fig4}}

\figcaption{
Many of the new FHLCs are bright enough that the SDSS spectra are useful
for astrophysical issues, as opposed to simply classification. Shown
here is a portion of the spectrum of SDSS~ J153732.2+004343, a previously
uncataloged $r^*\sim17.6$ object with strong $C_2$ Swan bands (not shown
in this segment) and a high proper motion, indicating a new dwarf C
star. The complete spectrum of this object appears in Figure 1. Note
the very strong NaD absorption: this is one of the cooler FHLCs in
our sample. The CaH bands are this strong only in dwarfs, and thus we
suggest that in this cooler subset of the FHLCs, these bands can serve as
the long-sought low-resolution spectroscopic luminosity discriminant,
eliminating the need for proper motion data. Note also the strong
BaII. 
\label{fig5}}

\figcaption{Results of Monte Carlo models which vary dwarf versus giant and
disk versus halo fraction of FHLCs, compared with the observed magnitude
({\it panel 'a'}), radial velocity ({\it panel~'b'}), and proper motion
({\it panel 'c'}) characteristics of the sample (excluding three known
extragalactic objects). In each case, the upper panel is the model,
and the lower panel illustrates one case which fits the data well: 15\%
halo giants, 50\% halo dwarfs, and 35\% disk dwarfs.
\label{fig6}}

\figcaption{The spectrum of SDSS~J171957.7+575005, a symbiotic carbon
star in the Draco dwarf galaxy, also known as Draco C1. In addition to
the $C_2$ Swan and CN bands, note the strong Balmer and HeI
$\lambda\lambda$ 5876, 6678, 7065 emission, and most especially the
extraordinary HeII $\lambda$4686 strength (HeII$\sim$H$\beta$). The
intense $H\alpha$ emission is truncated for convenience in scaling, but
the peak intensity is 700 units.
\label{fig7}}

\begin{figure}
\epsscale{1.0}
\plotone{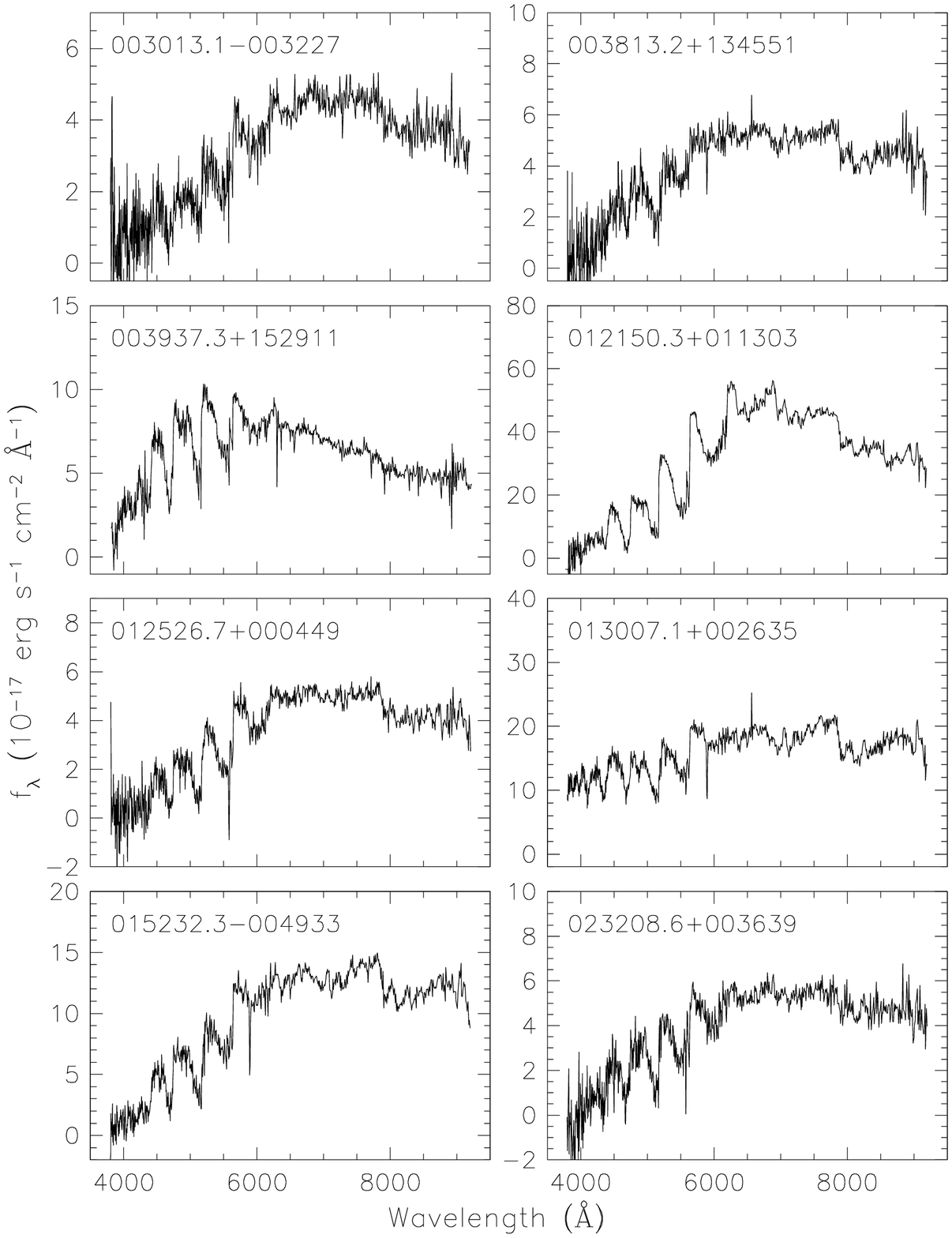}
\end{figure}

\begin{figure}
\epsscale{1.0}
\plotone{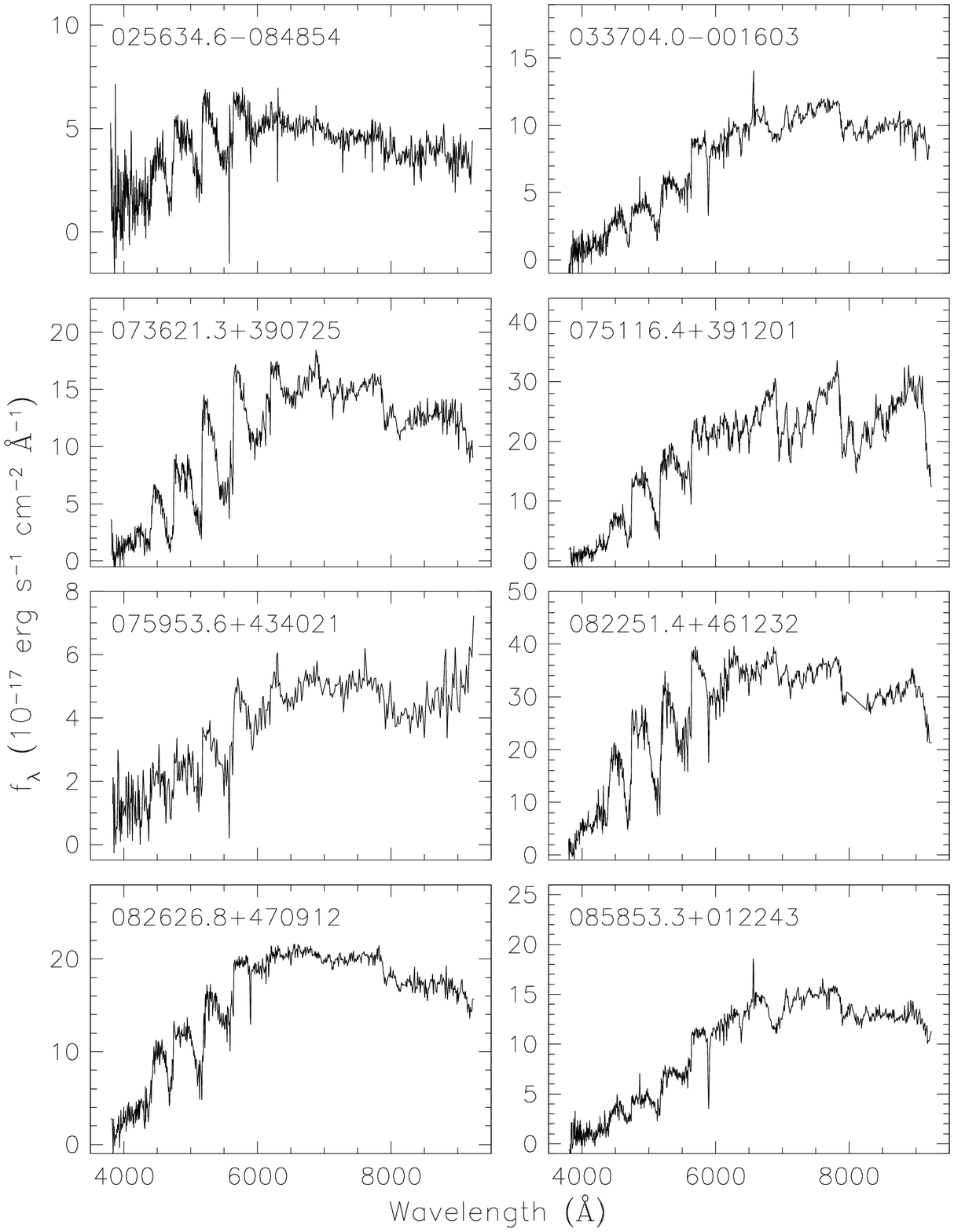}
\end{figure}

\begin{figure}
\epsscale{1.0}
\plotone{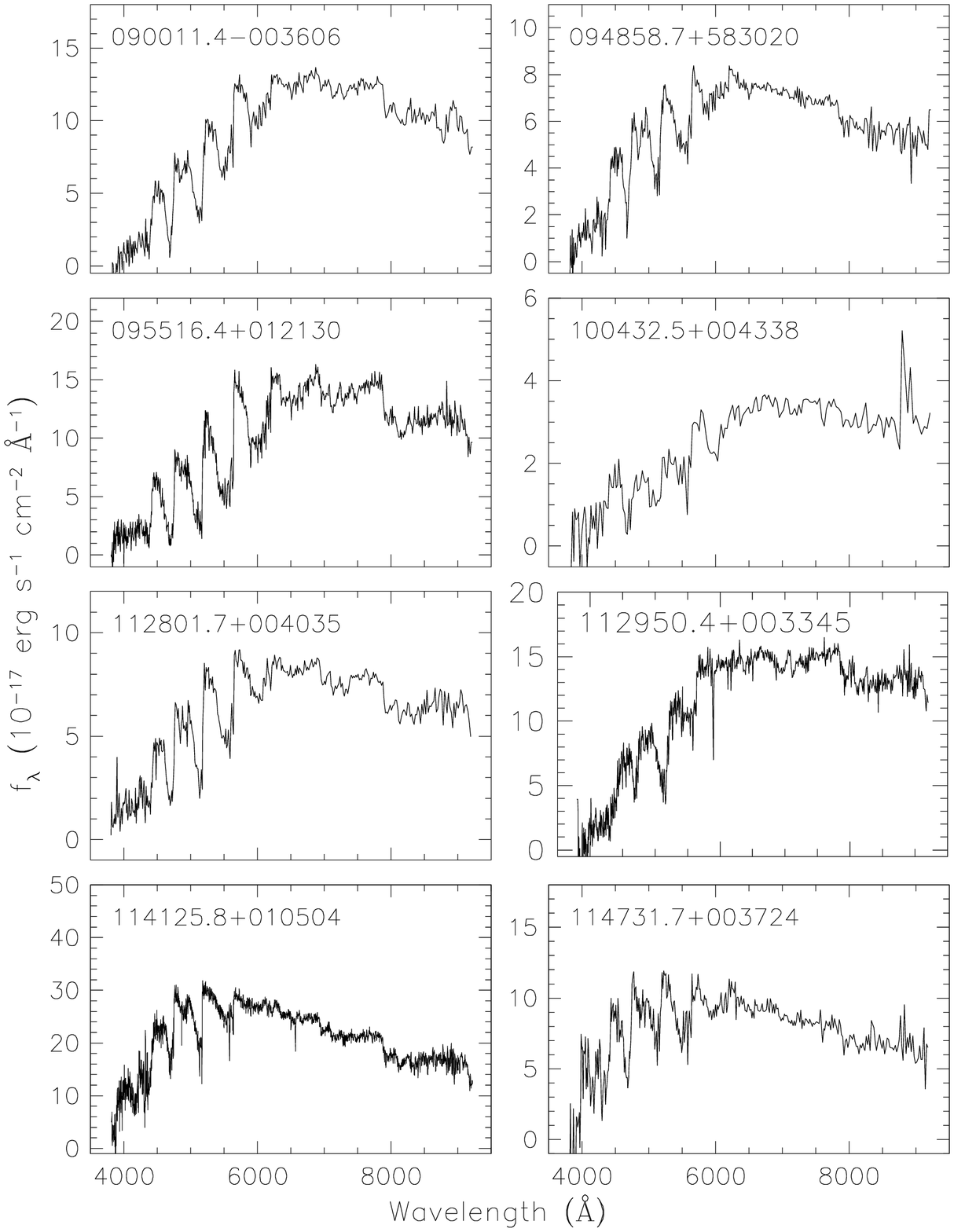}
\end{figure}

\begin{figure}
\epsscale{1.0}
\plotone{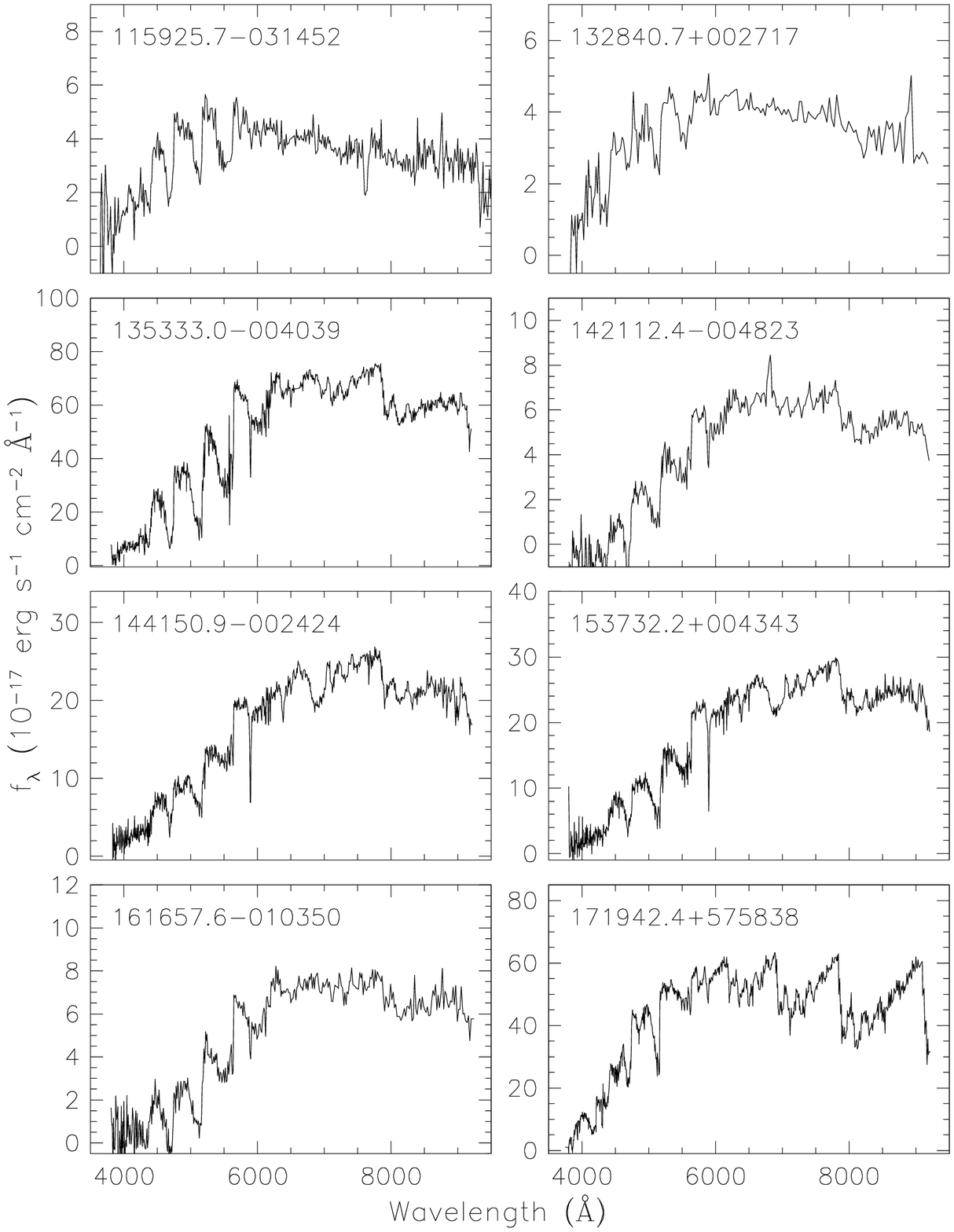}
\end{figure}

\begin{figure}
\epsscale{1.0}
\plotone{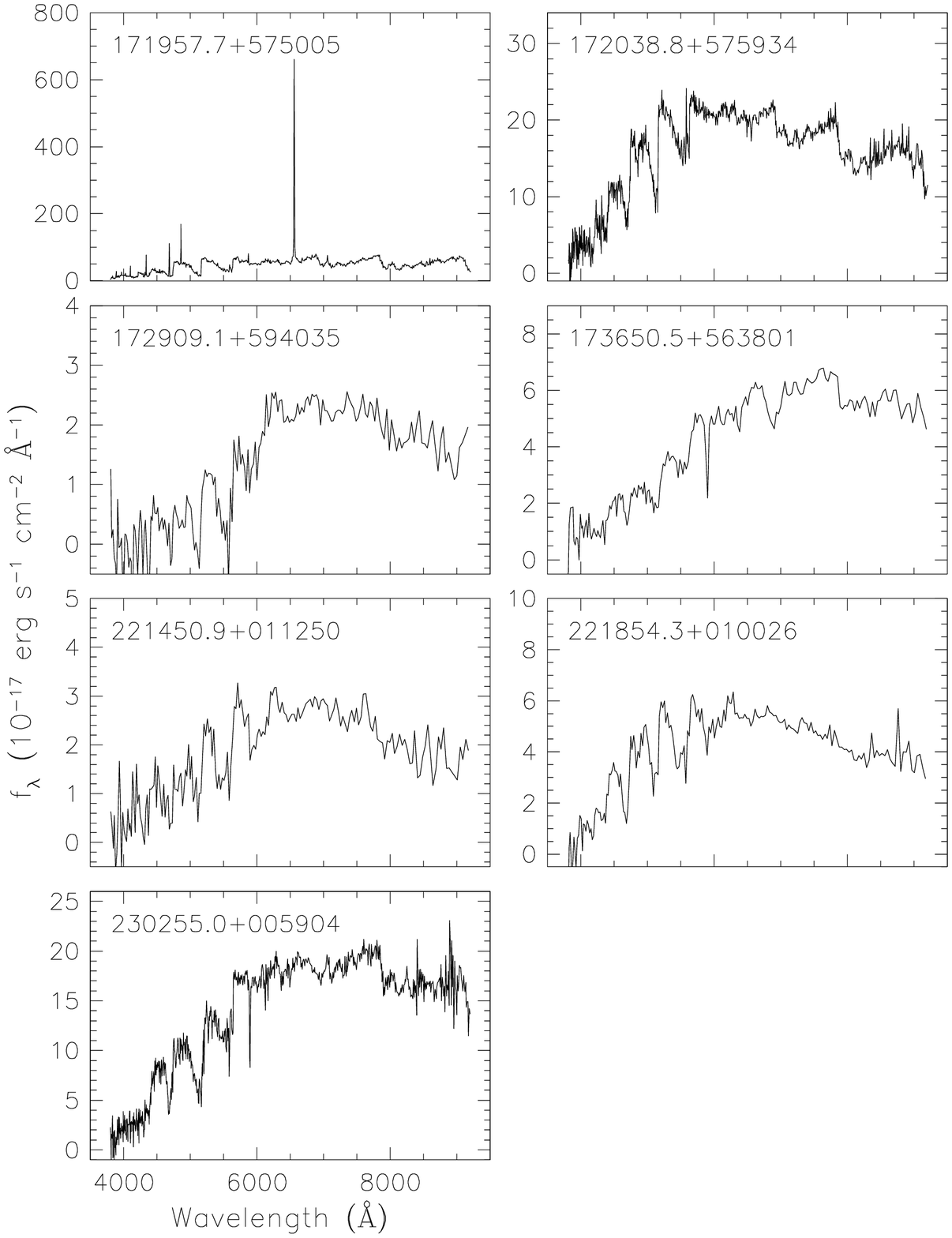}
\end{figure}

\begin{figure}
\epsscale{1.0}
\plotone{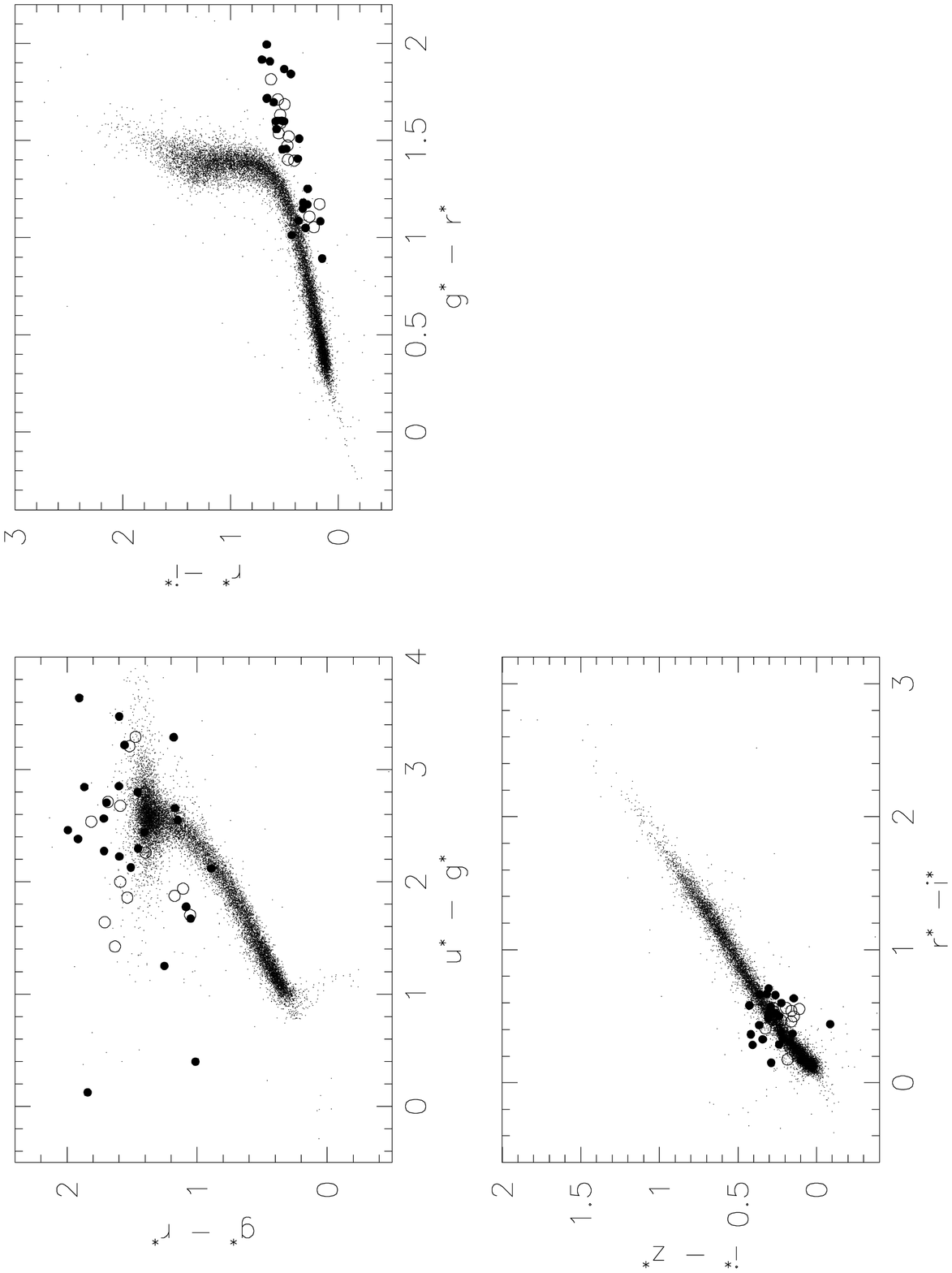}
\end{figure}

\begin{figure}
\epsscale{1.0}
\plotone{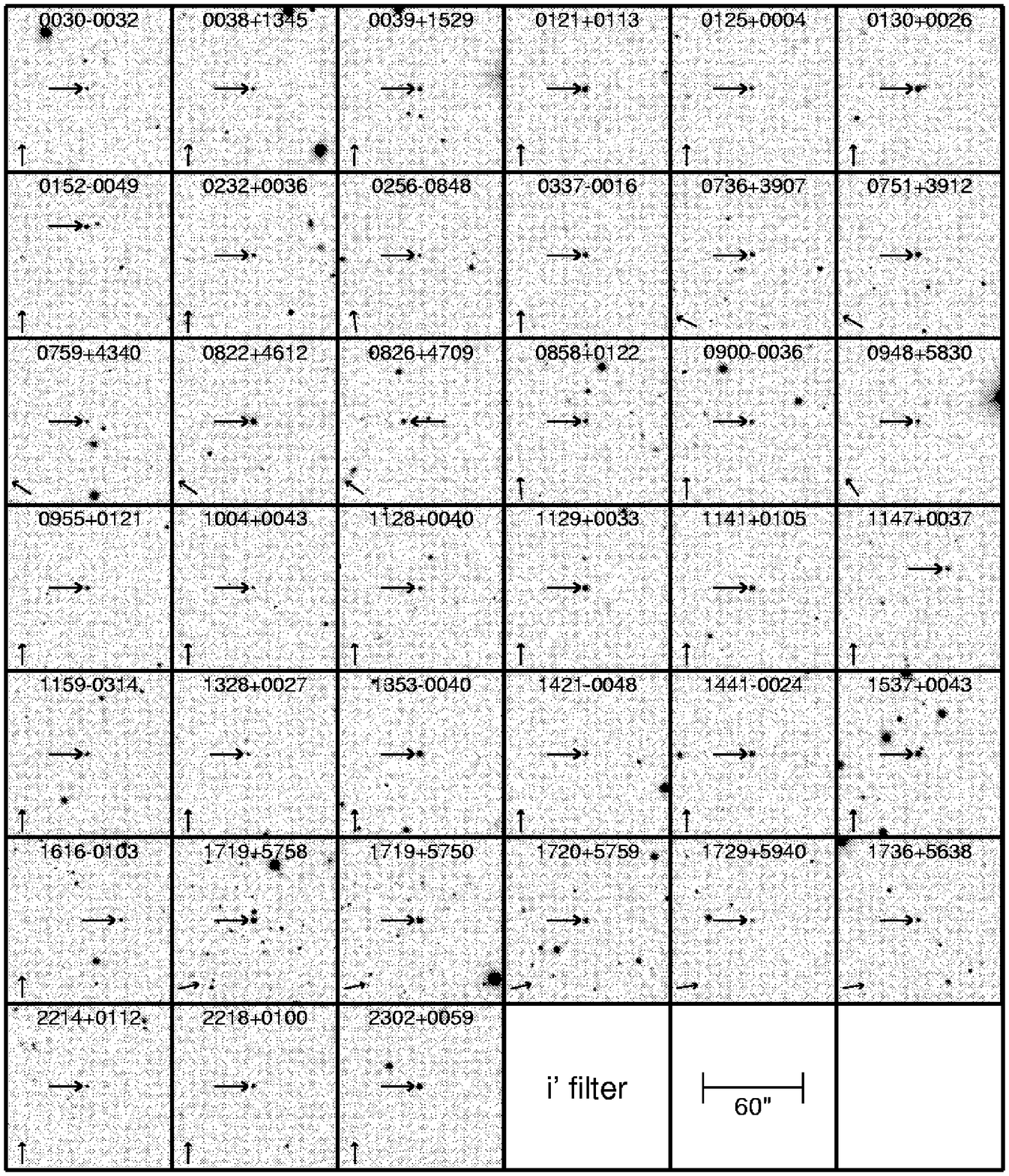}

\end{figure}

\begin{figure}
\epsscale{1.0}
\plotone{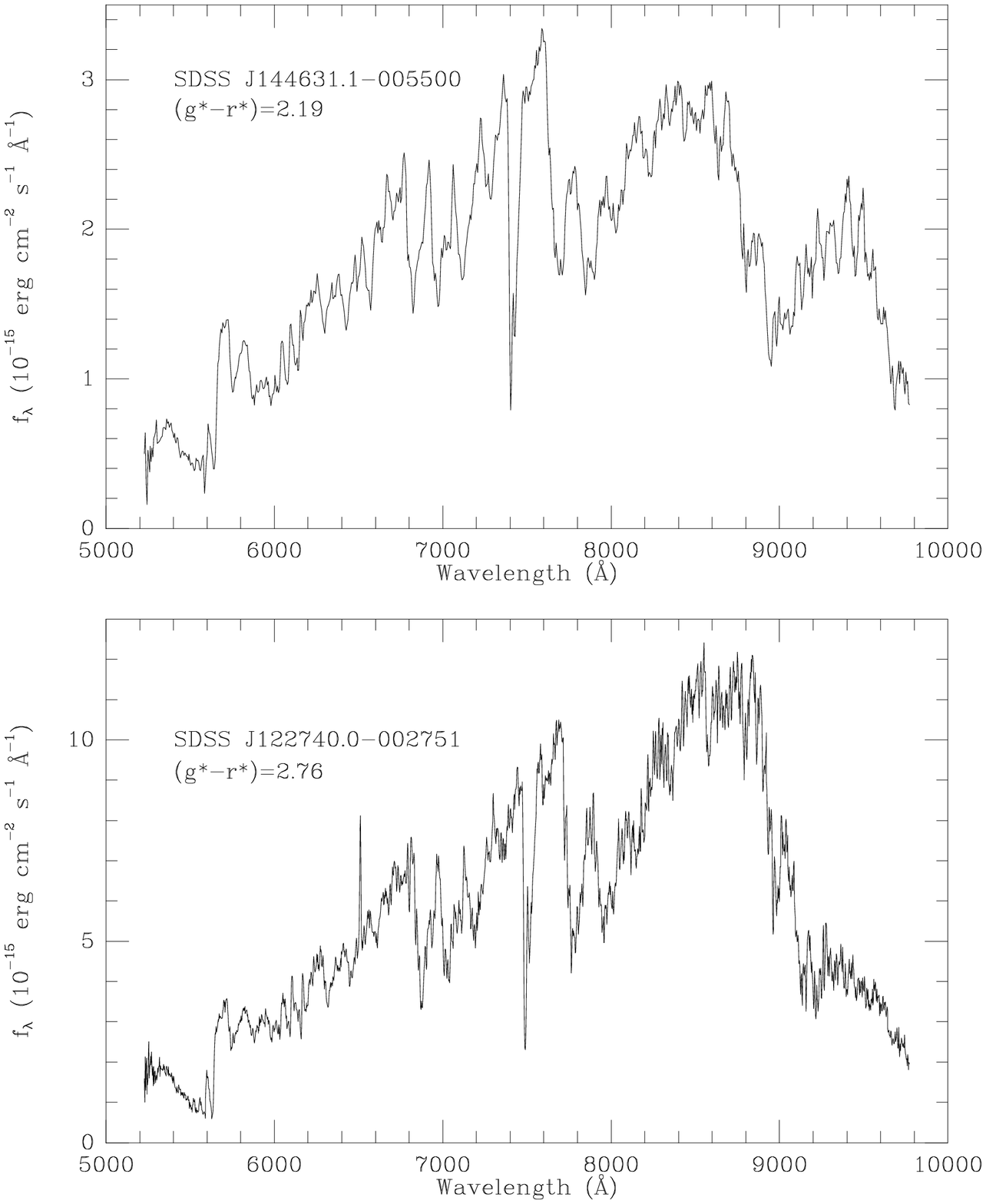}
\end{figure}

\begin{figure}
\epsscale{1.0}
\plotone{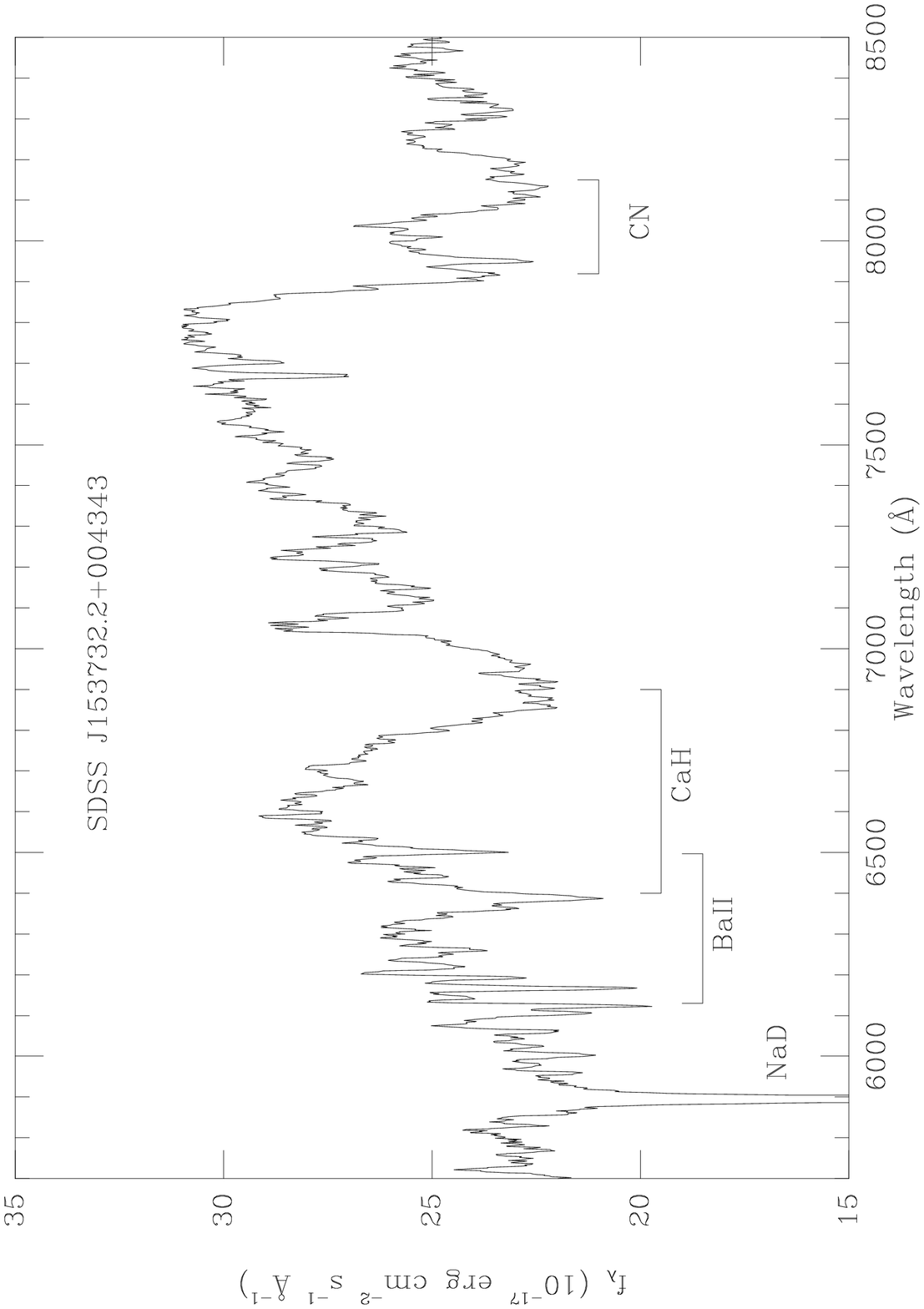}
\end{figure}

\begin{figure}
\epsscale{1.0}
\plotone{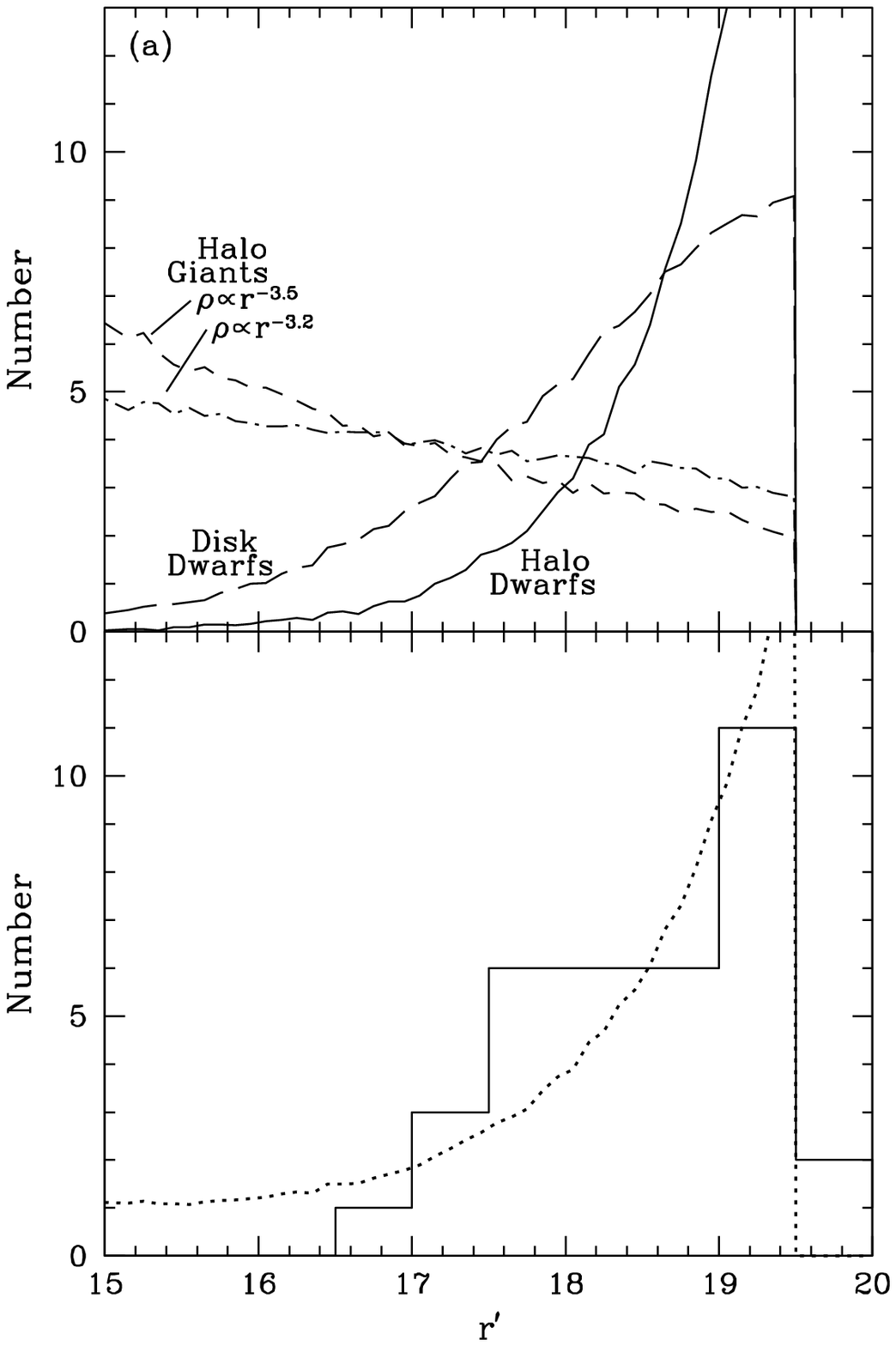}
\end{figure}

\begin{figure}
\epsscale{1.0}
\plotone{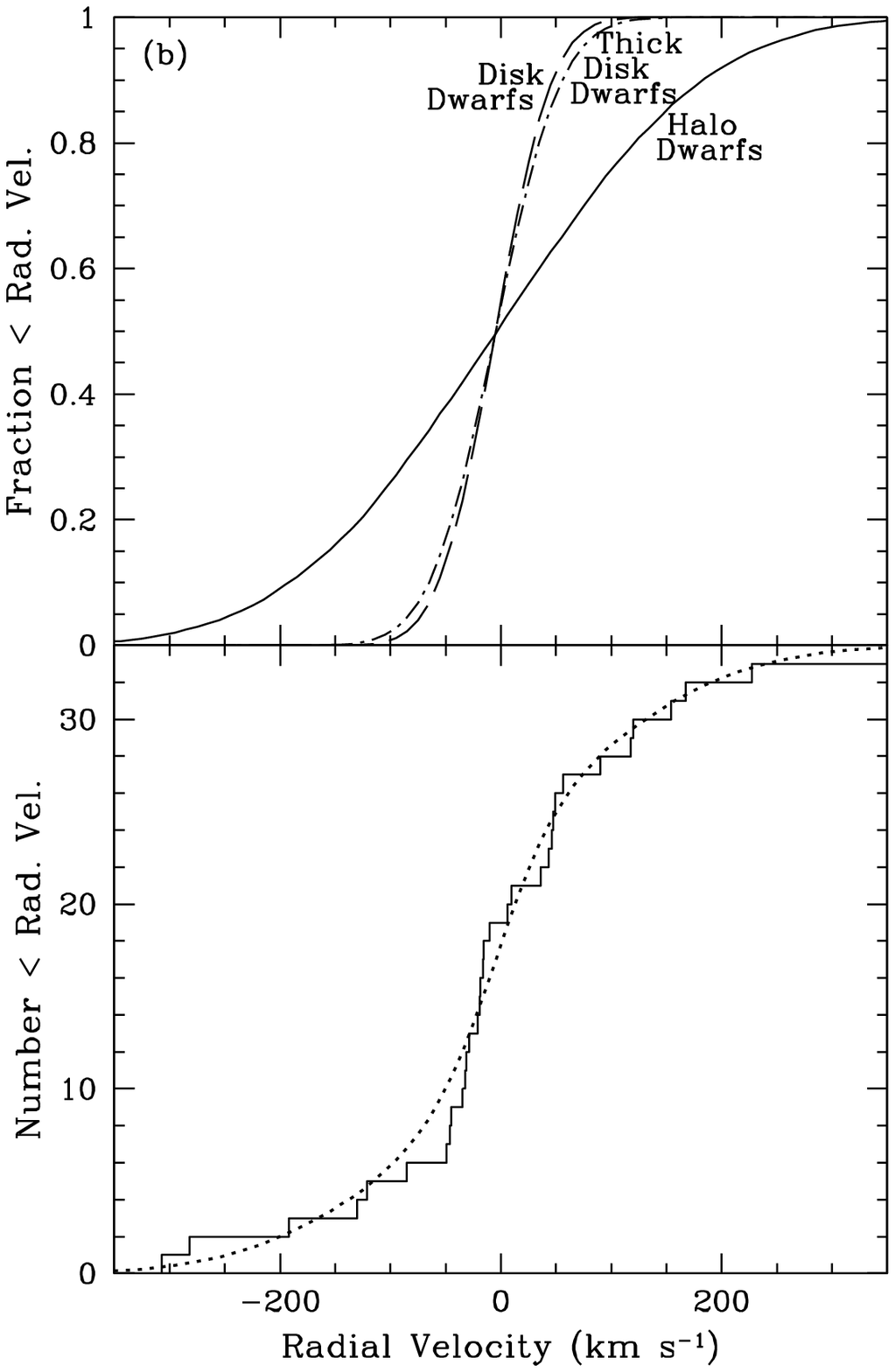}
\end{figure}

\begin{figure}
\epsscale{1.0}
\plotone{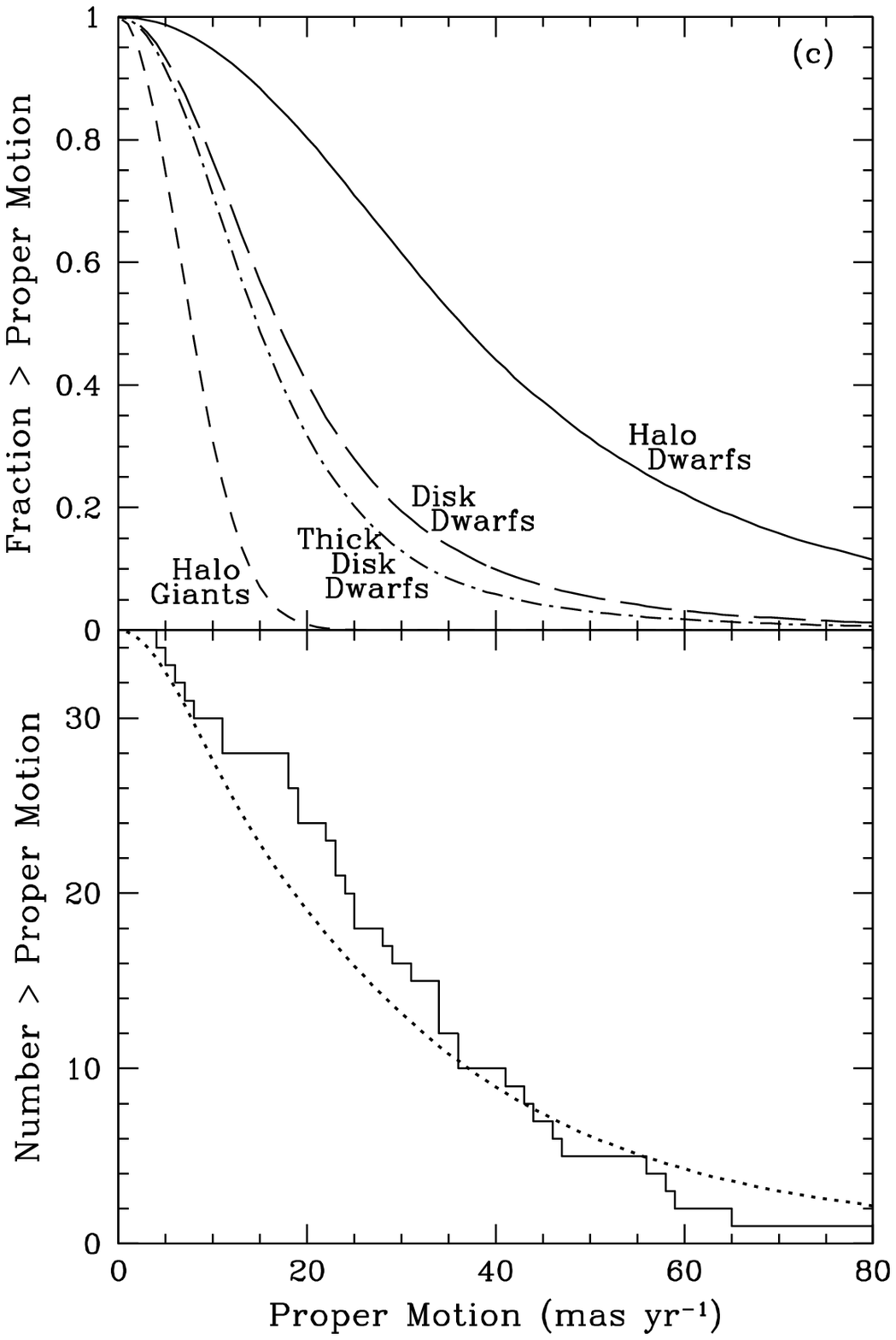}
\end{figure}

\begin{figure}
\epsscale{1.0}
\plotone{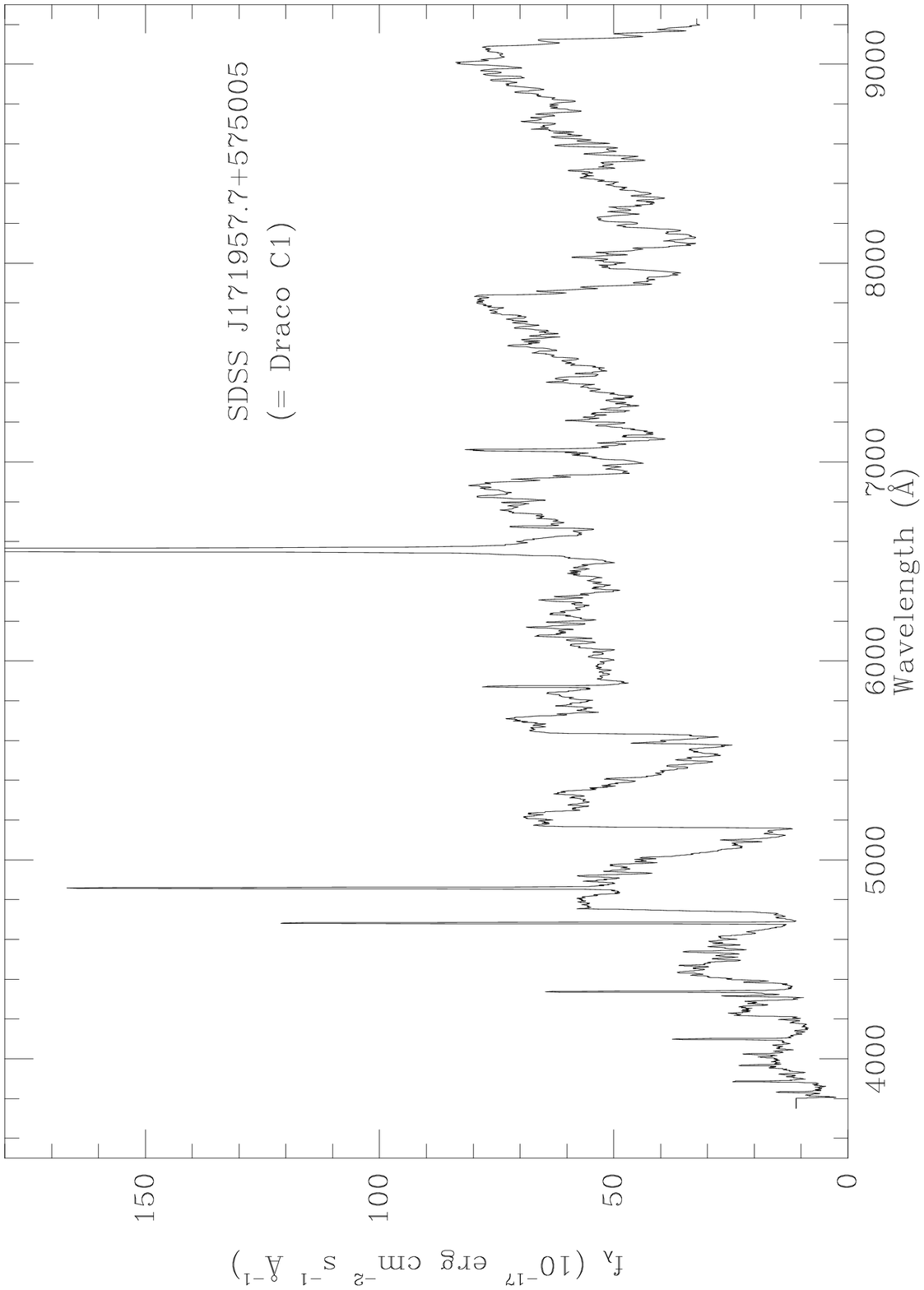}
\end{figure}


\begin{thebibliography}{}
\bibitem[Aaronson et al.~(1982)]{aar82} Aaronson, M., Liebert, J., \&
Stocke, J. 1982, \apj, 254, 507
\bibitem[Alksnis et al.~(2001)]{alk01} Alksnis, A., Balklavs, A., Dzervitis,~U., Eglitis,~I., 
Paupers,~O., \& Pundure,~I. 2001, Baltic Astr., 10, 1
\bibitem[Armandroff et al.~(1995)]{arm95} Armandroff, T.~E., Olszewski,
   E.~W., \& Pryor, C. 1995, \aj, 110, 2131
\bibitem[Baade \& Swope (1961)]{baa61} Baade, W., \& Swope, H. H. 1961, \aj,
    66, 300
\bibitem[Barnbaum et al. (1996)]{bar96} Barnbaum, C., Stone, R. P. S., \& Keenan, P. C. 1996, \apjs, 105, 419
\bibitem[Becker et al. (2001)]{bec01} Becker, R. H. et al. 2001, \aj,
   122, 2850
\bibitem[Bickert et al.~(1996)]{bic96} Bickert, K. F., Greiner, J., \&
  Stencel, R.~E. 1996, in Supersoft X-ray Sources, ed. J.~Greiner (Berlin: 
  Springer-Verlag), 225
\bibitem[Bond (1974)]{bon74} Bond, H. E. 1974, \apj, 194, 95
\bibitem[Bothun et al.~(1991)]{bot91} Bothun, G., Elias, J. H., MacAlpine, G.,
   Matthews, K., Mould, J. R., Neugebauer, G., \& Reid, I. N. 1991, \aj, 101,
   2220
\bibitem[Chiba \& Beers (2000)]{chi00} Chiba, M., \& Beers, T. C. 2000, \aj, 119, 2843
\bibitem[Christlieb et al.~(2001)]{chr01} Christlieb, N., Green, P. J.,   Wisotzki,~L., \& Reimers,~D. 2001, \aap, 375, 366
\bibitem[Cohen (1979)]{coh79} Cohen, M. 1979, \mnras, 186, 837
\bibitem[Cutri et al.~(1989)]{cut89} Cutri, R. M., Low, F. J., Kleinmann, S.
  G., Olszewski, E. W., Willner, S. P., Campbell, B., \& Gillett, F. C. 1989,
  \aj, 97, 866
\bibitem[Dahn et al.~(1977)]{dah77} Dahn, C. C., Liebert, J., Kron, R. G.,
   Spinrad, H., \& Hintzen, P.~M. 1977, \apj, 216, 757
\bibitem[Dearborn et al.~(1986)]{dea86} Dearborn, D. S. P., Liebert, J.,
   Aaronson, M., Dahn, C. C., Harrington, R., Mould, J., \& Greenstein, J. L.
   1986, \apj, 300, 314
\bibitem[deKool \& Green (1995)]{dek95} deKool, M., \& Green, P. J. 1995,
   \apj, 449, 236
\bibitem[Deutsch (1994)]{deu94} Deutsch, E.~W. 1994, \pasp, 106, 1134
\bibitem[Deutsch (1999)]{deu99} Deutsch, E.~W. 1999, \aj, 188, 1882
\bibitem[Fan et al. (2001)]{fan01} Fan, X. et al. 2001, \aj, 122, 2833
\bibitem[Finlator et al.~(2000)]{fin00} Finlator, K. et al., 2000, \aj, 120, 2615
\bibitem[Fujimoto et al.~(2000)]{fuj00} Fujimoto, M.~Y., Ikeda,~Y., \& Iben,~I. 2000, \apj, 529, L25
\bibitem[Fukugita et al.~(1996)]{fuk96} Fukugita, M., Ichikawa, T., Gunn, J.
  E., Doi, M., Shimasaku, K., \& Schneider, D. P. 1996, \aj, 111, 1748
\bibitem[Gass et al. (1988)]{gas88} Gass, H., Liebert, J., \&
   Wehrse,~R. 1988, \aap, 189, 194
\bibitem[Gigoyan et al. (2001)]{gig01} Gigoyan, K., Mauron, N., Azzopardi, M., 
Muratorio, G., \& Abrahamyan, H.~V. 2001, \aap, 371, 560
\bibitem[Gordon (1971)]{gor71} Gordon, C. P. 1971, \pasp, 83, 667
\bibitem[Green (2000)]{gre00} Green, P. J. 2000, in The Carbon Star Phenomenon,
  IAU Symp. 177, ed. R. F. Wing (Dordrecht: Kluwer), 27
\bibitem[Green \& Margon (1990)]{gre90} Green, P. J., and Margon, B. 1990,
   \pasp, 102, 1372
\bibitem[Green et al.~(1991)]{gre91} Green, P. J., Margon, B., \& MacConnell,
D. J. 1991, \apj, 380, L31
\bibitem[Green et al.~(1992)]{gre92} Green, P. J., Margon, B., Anderson, S.
  F., \& MacConnell, D. J. 1992, \apj, 400, 659
\bibitem[Green \& Margon (1994)]{gre94a} Green, P. J., \& Margon, B. 1994,
   \apj, 423, 723
\bibitem[Green et al.~(1994)]{gre94b} Green, P. J., Margon, B., Anderson, S.
  F., \& Cook, K. H. 1994, \apj, 434, 319
\bibitem[Gunn et al.~(1998)]{gun98} Gunn, J. E., et al. 1998, \aj, 116, 3040
\bibitem[Harris et al.~(1998)]{har98} Harris, H. C. et al., 1998, \apj, 502, 437
\bibitem[Heber et al.~(1993)]{heb93} Heber, U., Bade, N., Jordan, S., \&
  Voges, W. 1993, \aap, 267, L31
\bibitem[Hogg et al. (2001)]{hog01} Hogg, D.~W., Schlegel, D.~J., Finkbeiner, D.~P., \& Gunn, J.~E. 2001, \aj, 122, 2129
\bibitem[Ibata et al. (2001)]{iba01} Ibata, R., Lewis, G. F., Irwin, M.,
  Totten, E., \& Quinn, T. 2001, \apj, 551, 294
\bibitem[J\o rgensen et al. (1998)]{jor98} J\o rgensen, U. G., Borysow, A.,
  \& H\"ofner, S. 1997, in 1997 Pacific Rim Conference on Stellar
  Astrophysics, ed. K. L. Chan, K. S. Cheng, \& H. P. Singh (San Francisco:
  Astr. Soc. Pac.), 157
\bibitem[Joyce(1998)]{joy98} Joyce, R. R. 1998, \aj, 115, 2059
\bibitem[Keenan (1993)]{kee93} Keenan, P. C. 1993, \pasp, 105, 905
\bibitem[Kirkpatrick et al.~(1991)]{kir91} Kirkpatrick, J. D., Henry, T.~J., \&
   McCarthy, D.~W. 1991, \apjs, 77, 417
\bibitem[Krisciunas et al.~(1998)]{kri98} Krisciunas, K., Margon, B., \&
  Szkody,~P. 1998, \pasp, 110, 1342
\bibitem[Liebert et al. (2000)]{lie00} Liebert, J., Cutri, R. M., Nelson, B.,
  Kirkpatrick, J.~D., Gizis, J. E., \& Reid, I. N. 2000, \pasp, 112, 1315
\bibitem[Liebert et al. (1994)]{lie94} Liebert, J., Schmidt, G. D., Lesser,
  M., Stepanian, J. A., Lipovetsky, V. A., Chaffee, F. H, \& Foltz, C. B.
  1994, \apj, 421, 733
\bibitem[Longmore \& Allen (1977)]{lon77} Longmore, A. J., \& Allen, D. A. 1977, Ap. Lett., 18, 159
\bibitem[Lupton et al.~(2001)]{lup01} Lupton, R., Gunn, J. E., Ivezi\'c, Z., Knapp, G. R., Kent, S.. \& Yasuda, N. 2001, in  
ASP Conf. Ser. 238, Astronomical Data Analysis Software and
Systems X, ed. F.~R.~Harnden, Jr., F.~A.~Primini, and H. E. Payne (San
Francisco: Astr. Soc. Pac.), 269
\bibitem[Luyten (1979a)]{luy79a} Luyten, W. J. 1979a, NLTT Catalogue, Vol. II
  (Minneapolis: Univ. of Minnesota)
\bibitem[Luyten (1979b)]{luy79b} Luyten, W. J. 1979b, Proper Motion Survey
  with the Forty-Eight Inch Schmidt Telescope. Vol LII (Minneapolis: Univ. of
  Minnesota)
\bibitem[Luyten (1980)]{luy80} Luyten, W. J. 1980, NLTT Catalogue, Vol. III
   (Minneapolis: Univ. of Minnesota)
\bibitem[MacAlpine \& Williams (1981)]{mac81} MacAlpine, G. M., \& Williams,
   G. A. 1981, \apjs, 45, 113
\bibitem[Madau et al.~(1999)]{mad99} Madau, P., Haardt, F., \& Rees, M. J. 1999, \apj, 514, 648
\bibitem[Margon et al. (1984)]{mar84} Margon, B., Aaronson, M., Liebert, J.,
   \& Monet, D. 1984, \aj, 89, 274
\bibitem[Meusiner \& Brunzendorf (2001)]{meu01} Meusinger, H., \& Bruzendorf, J. 2001, IBVS 5035
\bibitem[Monet et al. (1998)]{mon98} Monet, D. G. et al., 1998, The USNO-A2.0 Catalog   (Washington: U.S. Naval Obs.)
\bibitem[Mould et al. (1985)]{mou85} Mould, J. R., Schneider, D. P., Gordon,
  G.~A., Aaronson, M., \& Liebert, J. W. 1985, \pasp, 97, 130
\bibitem[Munari (1991)]{mun91} Munari, U. 1991, \aap, 251, 103
\bibitem[M\"urset et al.~(1996)]{mur96} M\"urset, U., Jordan, S., \& Wolff,
  B. 1996, in Supersoft X-ray Sources, ed. J.~Greiner 
  (Berlin: Springer-Verlag), 251
\bibitem[Newberg et al. (2002)]{new02} Newberg, H. J. et al. 2002, \apj, 569, 
   245
\bibitem[Odenkirchen et al.~(2001)]{ode01} Odenkirchen, M. et al. 2001, \aj, 
   122, 2538
\bibitem[Olszewski et al.~(1995)]{ols95} Olszewski, E.~W., Aaronson, M., \& 
   Hill, J.M. 1995, \aj, 110, 2120
\bibitem[Piatek et al.~(2001)]{pia01} Piatek, S., Pryor, C., Armandroff, T. E.,   \&  Olszewski, E.~W. 2001, \aj, 121, 841
\bibitem[Pier et al. (2002)]{pie02} Pier, J. R., Munn, J. A., Hindsley, R. B.,
  Hennessy, G. S., Kent, S. M., Lupton, R. H., \& Ivezi\'c, Z. 2002,
  \aj, submitted
\bibitem[Rees (1998)]{ree98} Rees, M. J. 1998, \ssr, 84, 43
\bibitem[Reid et al. (1995)]{rei95} Reid, I. N., Hawley, S. L., \& Gizis, J. E. 1995, \aj, 110, 1838
\bibitem[Richards et al.~(2002)]{ric02} Richards, G. T. et al. 2002, \aj,
    123, 2945
\bibitem[Sanduleak \& Pesch (1988)]{san88} Sanduleak, N., \& Pesch, P. 1988,
  \apjs, 66, 387
\bibitem[Sanford (1949)]{san49} Sanford, R. F. 1944, \pasp, 61, 261
\bibitem[Schneider et al. (2002)]{sch02} Schneider, D. P., et al. 2002,
     \aj, 123, 567
\bibitem[Schulte-Ladbeck et al. (1988)]{sch88} Schulte-Ladbeck, R.E., MacConnell, D.~J., \& 
Zarate, N. 1988, in The Symbiotic Phenomenon, ed. J.~Mikolajewska et al (Doredrecht: Kluwer), 295
\bibitem[Skrutskie et al. (1997)]{skr97} Skrutskie, M. F., et al. 1997, in
   The Impact of Large Scale Near-IR Sky Surveys, ed. F.~Garzon et al.
   (Dordrecht: Kluwer), 187
\bibitem[Smith et al. (2002)]{smi02} Smith, J. A., et al. 2002, \aj, 123, 2121
\bibitem[Stoughton et al. (2002)]{sto02} Stoughton, C. et al. 2002,
   \aj, 123, 485
\bibitem[Soyano \& Maehara (1999)]{soy99} Soyano, T., \& Maehara,~H. 1999, Pub. Nat. Astr. Obs. Japan, 5, 149
\bibitem[Totten \& Irwin (1998)]{tot98} Totten, E. J., \& Irwin, M. J. 1998,
   \mnras, 294, 1
\bibitem[Totten et al. (2000)]{tot00} Totten, E. J., Irwin,~M.
   J., \& Whitlock, P. A. 2000, \mnras, 314, 630
\bibitem[Wing \& J\o rgensen (1996)]{win96} Wing, R. F., \& J\o rgensen, U. G.
  1996, \baas, 28, 1382
\bibitem[York et al.~(2000)]{yor00} York, D. G. et al., 2000, \aj, 120, 1579

\end{thebibliography}
\end{document}